\documentclass[12pt,a4paper]{iopart}
\usepackage[english]{babel}

\usepackage{iopams}
\usepackage{verbatim}
\usepackage{color}

\usepackage{graphicx}

\usepackage[colorlinks,citecolor=blue,linkcolor=blue,urlcolor=blue]{hyperref}

\def\iotabar{\lower3pt\hbox{$\mathchar'26$}\mkern-7mu\iota}
\newcommand {\aplt}{\ {\raise-.5ex\hbox{$\buildrel<\over\sim$}}\ }
\newcommand{\dd}{\mbox{d}}

\newcommand{\spe}{{\sigma}}
\newcommand{\lw}{{\rm lw}}
\newcommand{\sw}{{\rm sw}}
\newcommand{\eq}[1]{(\ref{#1})}
\newcommand{\bun}{\hat{\mathbf{b}}}
\newcommand{\eun}{\hat{\mathbf{e}}}

\newcommand{\phiwig}{\widetilde{\phi}}
\newcommand{\Phiwig}{\widetilde{\Phi}}
\newcommand{\boldr}{\mathbf{r}}
\newcommand{\bv}{\mathbf{v}}
\newcommand{\bA}{\mathbf{A}}
\newcommand{\bR}{\mathbf{R}}

\newcommand{\bJ}{\mathbf{J}}

\newcommand{\bB}{\mathbf{B}}

\newcommand{\bZ}{\mathbf{Z}}
\newcommand{\bK}{\mathbf{K}}

\newcommand{\bX}{\mathbf{X}}
\newcommand{\bV}{\mathbf{V}}

\newcommand{\matrixtop}[1]{\buildrel\leftrightarrow\over{#1}}
\newcommand{\matI}{\matrixtop{\mathbf{I}}}

\newcommand{\matP}{\matrixtop{\mathbf{P}}}
\newcommand{\matW}{\matrixtop{\mathbf{W}}}
\newcommand{\matM}{\matrixtop{\mathbf{M}}}

\newcommand{\dotcross}{ \raise 0.65ex\hbox{${\scriptstyle {{_{\displaystyle \cdot}}\atop\times}}$} }
\newcommand{\crossdot}{ \raise 0.5ex\hbox{${\scriptstyle {{_\times}\atop{\displaystyle \cdot}}}$} }
\newcommand{\rhobf}{\mbox{\boldmath$\rho$}}

\newcommand{\kappabf}{\mbox{\boldmath$\kappa$}}
\newcommand{\zetabf}{\mbox{\boldmath$\zeta$}}

\newcommand{\zun}{\hat{\zetabf}}

\newcommand{\cT}{{\cal T}}
\newcommand{\modTinv}{{\mathbb T_{\spe,0}}}
\newcommand{\modTinvprime}{{\mathbb T_{\spe',0}}}

\newcommand{\sumsig}{ \raise -1.3ex\hbox{${{\displaystyle \sum}\atop{\scriptstyle \sigma}}$} }
\newcounter{appnumb}

\begin{document}

\title[Radial transport of toroidal angular momentum in
tokamaks]{Radial transport of toroidal angular momentum in
tokamaks}
\author{Iv\'an Calvo$^{1}$} \eads{\mailto{ivan.calvo@ciemat.es}}
\author{Felix I Parra$^{2,3}$}
\eads{\mailto{felix.parradiaz@physics.ox.ac.uk}}

\vspace{0.5cm}

\address{$^1$Laboratorio Nacional de Fusi\'on, CIEMAT, 28040 Madrid, Spain}
\address{$^2$Rudolf Peierls Centre for Theoretical Physics, 
University of Oxford, Oxford, OX1 3NP, UK}
\address{$^3$Culham Centre for Fusion Energy, Abingdon, OX14 3DB, UK}

%\submitto{\PPCF}

\vskip 1cm

%%{\large
%%\begin{center}
%%\today
%%\end{center}
%%}

\begin{abstract}
  The radial flux of toroidal angular momentum is needed to determine
  tokamak intrinsic rotation profiles. Its computation requires
  knowledge of the gyrokinetic distribution functions and turbulent
  electrostatic potential to second-order in $\epsilon = \rho/L$,
  where $\rho$ is the ion Larmor radius and $L$ is the variation
  length of the magnetic field. In this article, a complete set of
  equations to calculate the radial transport of toroidal angular
  momentum in any tokamak is presented. In particular, the
  $O(\epsilon^2)$ equations for the turbulent components of the
  distribution functions and electrostatic potential are given for the
  first time without assuming that the poloidal magnetic field over
  the magnetic field strength is small.
\end{abstract}

\maketitle

%%%%%%%%%%%%%%%%%%%%%%%%%%%%%%%%%%%%%%%%%%%%%%%%%%%%%%%%%%%%%%%
\section{Introduction}
\label{sec:Introduction}
%%%%%%%%%%%%%%%%%%%%%%%%%%%%%%%%%%%%%%%%%%%%%%%%%%%%%%%%%%%%%%%

Intrinsic rotation is  {the toroidal} rotation that tokamak
plasmas can achieve without the need of external momentum input. There
is a number of reasons why intrinsic rotation has become an active
research field in the last years~\cite{ParraBarnesCalvoCatto2012,
  Parra2015}. First, fast toroidal rotation seems to improve
confinement~\cite{Mantica09} and it quenches MHD instabilities
  \cite{devries96}. Second, it is unlikely that rotation in the core
of large and dense tokamaks can be produced by means of neutral beam
momentum injection~\cite{Greenwald05, deGrassie07}. Therefore, in
tokamaks like ITER or larger, the only rotation available will be
intrinsic. Third, Parra and Catto showed that standard code
implementations of the gyrokinetic equations cannot determine
the intrinsic rotation profile; they also pointed out that the correct
equations had not been derived yet~\cite{parra08, parra09b, parra10b,
  parra10c}. We explain the problem in more detail in what follows.

Gyrokinetic theory~\cite{catto78} is the appropriate framework to
study microturbulence in fusion plasmas; that is, turbulence on the
ion Larmor radius scale, $\rho$, with characteristic frequencies of
the order of $c_s/L$, where $c_s$ is the sound speed and $L$ is the
typical variation length of the magnetic field. The theory is
formulated as an asymptotic expansion in $\epsilon := \rho/L \ll 1$,
in which the gyromotion degree of freedom is eliminated order by order
in $\epsilon$, whereas non-zero Larmor radius effects are
retained. The specific orderings assumed in electrostatic gyrokinetics
are explained in Section \ref{sec:orderingsandassumptions}. They lead
to a consistent set of equations, as proven, in particular, by the
fact that gyrokinetic simulations~\cite{dorland00, dannert05, candy03,
  chen03, peeters09} converge as long as the size of the simulation
domain exceeds several gyroradii. Recently, it has explicitly been
checked that the fluctuation spectrum reaches its maximum at
wavelengths of the order of $\rho$~\cite{barnes11}. Finally, dedicated
experiments have confirmed that the scaling of the turbulence
characteristics with the gyroradius are reproduced by gyrokinetic
theory~\cite{mckee01}. An introductory survey of gyrokinetics can be
found in \cite{krommes2012}.

The so-called low-flow ordering, in which the plasma velocity is
assumed to be of order $\epsilon c_s$, is probably the relevant one in
the core of large and dense tokamaks. The reason is that sonic speeds
are typically reached by neutral beam momentum injection, and we have
already mentioned that its efficiency is reduced in large machines. In
the low-flow regime, and in up-down symmetric
tokamaks~\cite{parra11c}, Parra and Catto proved that finding the
intrinsic rotation profile requires the gyrokinetic equations to
$O(\epsilon^2)$. Determining the toroidal rotation is equivalent to
determining the long-wavelength radial electric field, and viceversa,
because the neoclassical relation between the two
quantities~\cite{hinton76} still holds in the presence of
microturbulence. Then, the difficulty of computing the toroidal
rotation can be traced back to the intrinsic ambipolarity of the
tokamak; that is, to the fact that to lowest order, and even in the
presence of turbulence, the flux-surface-averaged ion and electron
current densities cancel each other for any value of the
long-wavelength radial electric field\footnote{ { {The intrinsic ambipolarity of the tokamak is a result of its axisymmetry.} Therefore, the
    problem disappears for sufficiently rippled tokamaks or
    stellarators far from quasisymmetry~\cite{Helander08,
      CalvoParraVelascoAlonso13, CalvoParraAlonsoVelasco14,
      CalvoParraVelascoAlonso15}.}}. However, gyrokinetic codes only
have the equations implemented to such lowest order.

As advanced above, Parra and Catto~\cite{parra10a} arrived at a
formula for the radial flux of toroidal angular momentum that is
written in terms of velocity integrals involving up to $O(\epsilon^2)$
pieces of the ion distribution functions and electrostatic
potential. Hence, the Fokker-Planck and quasineutrality equations have
to be solved to $O(\epsilon^2)$ to evaluate the radial flux of
toroidal angular momentum. The set of equations appropriate for the
limit $B/B_p \gg 1$ has been given in references~\cite{parra10a,
  parra11NF, Parra2015}. Here, $B_p$ is the magnitude of the
poloidal magnetic field. However, the equations valid for a general
tokamak had not been calculated.

In reference \cite{CalvoParra2012}, by employing the results of
\cite{ParraCalvo2011}, the first step to compute the toroidal angular
momentum flux without using the approximation $B/B_p \gg 1$ has been
taken. In order to be more specific about what has been done and what
remained to be done, we recall that in gyrokinetics the fields of the
theory, i.e. the distribution functions and the electrostatic
potential, are naturally decomposed as the sum of their
long-wavelength and short-wavelength components. The short-wavelength
components correspond to the turbulence scales and the long-wavelength
ones to the macroscopic scales (see Section
\ref{sec:orderingsandassumptions} for more details on the
decomposition). Both components of the distribution functions to
$O(\epsilon^2)$ enter the momentum transport equations. As for the
electrostatic potential, the turbulent component to $O(\epsilon^2)$ is
needed, but the long-wavelength component is required only to
$O(\epsilon)$. In \cite{CalvoParra2012}, the long-wavelength
gyrokinetic equations up to $O(\epsilon^2)$ have been derived. In this
paper, we compute the short-wavelength equations and then write
explicitly all the equations that, when implemented in a global
$\delta f$ code, give the toroidal angular momentum flux in any
tokamak to the order needed to calculate intrinsic rotation.

In Section \ref{sec:orderingsandassumptions}, the orderings assumed in
low-flow electrostatic gyrokinetics are explained. Section
\ref{sec:radialFluxTorAngMom} summarizes the procedure employed by
Parra and Catto to reach their expression for the radial flux of
toroidal angular momentum. In Section \ref{sec:GyrokineticExpansion}
we show that the evaluation of such an expression requires the
long-wavelength and short-wavelength components of the distribution
functions to $O(\epsilon^2)$, the long-wavelength component of the
electrostatic potential to $O(\epsilon)$, and the short-wavelength
component of the electrostatic potential to $O(\epsilon^2)$. In
Section \ref{sec:SWequations} the equations for the short-wavelength
components are provided. In subsection
\ref{sec:FPandQuasineutSWfirstOrder}, we simply recall the well-known
equations for the $O(\epsilon)$ turbulent components. In subsections
\ref{sec:shortWaveFokkerPlanckEq} and \ref{sec:shortWavePoissonEq},
the equations for the $O(\epsilon^2)$ turbulent components of the
distribution functions and electrostatic potential are derived for the
first time. Section \ref{sec:LongWave} contains the results of
reference \cite{CalvoParra2012} that are needed to formulate a
complete set of equations for toroidal angular momentum transport in any
tokamak. These are the equations for the long-wavelength components of
the distribution functions up to $O(\epsilon^2)$, the equation for the
long-wavelength component of the electrostatic potential to
$O(\epsilon)$, the transport equation for the density of each species,
and the transport equation for the total energy. In Section
\ref{sec:discussion} we summarize the results and list the
equations of the main text that should be implemented in a code to
determine the toroidal rotation profile.

%%%%%%%%%%%%%%%%%%%%%%%%%%%%%%%%%%%%%%%%%%%%%%%%%%%%%%%%%%%%%%%
\section{Orderings and assumptions}
\label{sec:orderingsandassumptions}
%%%%%%%%%%%%%%%%%%%%%%%%%%%%%%%%%%%%%%%%%%%%%%%%%%%%%%%%%%%%%%%

The starting point in the derivation of the electrostatic gyrokinetic
equations consists of the Fokker-Planck equation for each species
$\spe$,
\begin{eqnarray}\label{eq:FPinitial}
\fl\partial_t f_\spe + \bv\cdot\nabla_\boldr f_\spe
+\frac{Z_\spe e}{m_\spe}\left(-\nabla_\boldr\varphi +
\frac{1}{c}\, \bv\times\bB\right)\cdot\nabla_\bv f_\spe =\nonumber\\[5pt]
\fl\hspace{1cm}
\sum_{\spe'}C_{\spe \spe'}[f_\spe,f_{\spe'}] + S_\spe,
\end{eqnarray}
and the quasineutrality equation,
\begin{eqnarray}\label{eq:quasineutrality}
\sum_\spe Z_\spe e\int \,
f_\spe(\boldr,\bv,t)\dd ^3 v
=
0
.
\end{eqnarray}
The fields of the theory are the phase-space probability distributions
$f_\spe(\boldr,\bv,t)$ and the electrostatic potential
$\varphi(\boldr, t)$. The magnetic field
$\bB(\boldr)=\nabla_\boldr\times\bA(\boldr)$ does not vary in time,
$c$ is the speed of light, $e$ the charge of the proton, and $Z_\spe
e$ and $m_\spe$ are the charge and the mass of species $\spe$. The
Landau collision operator between species $\spe$ and $\spe'$ reads
\begin{eqnarray}\label{eq:collisionoperator}
C_{\spe \spe'}[f_\spe,f_{\spe'}](\boldr,\bv,t) =
\nonumber\\[5pt]
\hspace{1cm}
\frac{\gamma_{\spe\spe'}}{m_\sigma}
\nabla_\bv\cdot
\int
\matW(\bv-\bv')
\cdot
\Bigg(
\frac{1}{m_\spe}f_{\spe'}(\boldr,\bv',t)\nabla_\bv f_\spe(\boldr,\bv,t)
\nonumber\\[5pt]
\hspace{1cm}
-
\frac{1}{m_{\spe'}}
f_{\spe}(\boldr,\bv,t)\nabla_{\bv'} f_{\spe'}(\boldr,\bv',t)
\Bigg)
\dd^3v',
\end{eqnarray}
where
\begin{equation}
\gamma_{\spe\spe'}:= 2\pi Z_\spe^2 Z_{\spe'}^2 e^4 \ln\Lambda,
\end{equation}
\begin{equation}
{\bf \matW}({\bold w})
 := \frac{|{\bold w}|^2{\matI}-{\bold w}{\bold w}}{|{\bold w}|^3},
\end{equation}
$\ln\Lambda$ is the Coulomb logarithm, and $\matI$ is the identity
matrix. Finally, $S_\spe$ is a source term that represents heating and
fuelling in the tokamak.  {Our equation does not
  explicitly include the transformer electric field because we assume
  that the electric field is purely electrostatic, but the effect of
  the transformer electric field can be recovered with suitable
  choices of $S_\spe$.}

Gyrokinetic theory relies on the existence of very different spatial
and time scales when the plasma is strongly magnetized. The small
spatial scale is given by the gyroradius of species $\spe$, $\rho_\spe
= v_{t\spe}/\Omega_\spe$, and the large spatial scale is given by $L
\sim |\nabla_\boldr B /B|^{-1}$, where $B = |\bB|$. Here, $v_{t\spe} =
\sqrt{T_{0}/m_\spe}$ is the thermal velocity of species $\spe$, where
$T_{0}$ is a typical value of the temperature, and $\Omega_\spe =
Z_\spe e B_0 / (m_\spe c)$, the gyrofrequency, is defined in terms of
a typical value of the magnetic field, $B_0$. Strong magnetization
implies that $\epsilon_\spe := \rho_\spe / L \ll 1$; that is, the
charged particle gyrates around a magnetic field line separating from
it only a small distance of the order of the gyroradius. There is also
a large separation in time scales. The fast time scale is the
gyrofrequency $\Omega_\spe$ and the slow time scale is $v_{t\spe}/L$
(relevant turbulence frequencies satisfy $\omega \sim v_{t\spe}/L
$). In gyrokinetic theory, one eliminates the degree of freedom
corresponding to the gyromotion by averaging out frequencies of the
order of $\Omega_\spe$, while non-zero Larmor radius effects are
kept. Even if we often use $\epsilon_\spe$ for convenience, we need a
species-independent expansion parameter. We choose $\epsilon_s :=
\rho_s /L$, where $\rho_s = c_s/\Omega_i$ is the sound gyroradius and
$c_s=\sqrt{T_{0}/m_i}$ is the sound speed. We assume that the
dominant ion species, denoted by the subindex $i$, is singly charged;
i.e. $Z_i = 1$. Gyrokinetics is then formulated as an asymptotic
expansion in $\epsilon_s$. Observe that the relation between
$\epsilon_\spe$ and $\epsilon_s$ is
$\epsilon_s=Z_\spe\tau_\spe\epsilon_\spe$, with
\begin{equation}
\tau_\spe = \frac{v_{t\spe}}{c_s} =
\sqrt{\frac{m_i}{m_\spe }}\, .
\end{equation}

The presence of disparate time and space scales makes it useful to
write the fields of the theory as the sum of two components, one
associated to the microturbulence scales (the short-wavelength
component) and the other to the macroscopic scale (the long-wavelength
component). To properly define these components, we need a set of flux
coordinates $\{\psi,\Theta,\zeta\}$, where $\psi$ is the poloidal
magnetic flux, $\Theta$ is the poloidal angle, and $\zeta$ is the
toroidal angle. The magnetic field of the tokamak is written as
\begin{equation}
\bB = I(\psi)\nabla_\boldr\zeta + \nabla_\boldr\zeta\times\nabla_\boldr\psi.
\end{equation}
The coordinate $\zeta$ is chosen to be the standard toroidal angle in
the cylindrical coordinates whose axis is the tokamak symmetry
axis. In particular, $|\nabla_\boldr\zeta| = R^{-1}$, where $R$ is the
distance to the symmetry axis.

The axisymmetric long-wavelength
component of a function $g(\psi,\Theta,\zeta)$ is defined by
\begin{equation}\label{eq:deflwcomponent}
g^\lw
= \frac{1}{2\pi\Delta t \Delta \psi \Delta\Theta}
\int_{\Delta t}\dd t
\int_{\Delta \psi}\dd \psi
\int_{\Delta \Theta}\dd \Theta\int_0^{2\pi}\dd\zeta\, g,
\end{equation}
where $\epsilon_s\ll \Delta\psi/\psi \ll 1$, $\epsilon_s\ll
\Delta\Theta \ll 1$, and $L/c_s\ll\Delta t\ll \tau_E$. Here,
\begin{equation}
\tau_E := \frac{L}{\epsilon_s^{2}c_s}
\end{equation}
is the transport time scale (i.e. the time
scale of variation of the profiles) and $L/c_s$ is the time scale of
the turbulence. Then, the short-wavelength component is defined as
\begin{eqnarray}
g^\sw &:= g - g^\lw.
\end{eqnarray}
For any two functions $g(\boldr,t)$ and $h(\boldr,t)$ we have
\begin{eqnarray}
&&\left[g^\lw\right]^\lw = g^\lw,\nonumber\\[5pt]
&&\left[g^\sw\right]^\lw = 0,\nonumber\\[5pt]
&&\left[g h\right]^\lw = g^\lw h^\lw + \left[g^\sw h^\sw\right]^\lw.
\end{eqnarray}
 {The decomposition into a short-wavelength and a
  long-wavelength component is not unambiguously defined  {because the
  integration limits of \eq{eq:deflwcomponent} are not specified. For the most part, this ambiguity is not a problem because the long- and short-wavelength pieces can be thought of as defined by the equations they satisfy. However, to implement some of the equations in this article, one needs to choose a time interval and a space volume over which the average $[\ldots]^\lw$ is taken. For example, in equations \eq{eq:sworder1distfunction}, \eq{eq:gyroaveragedF2sw}, \eq{eq:quasineutralityOrder2sw} and \eq{eq:eqH2sigma}, we need to take the average $[\ldots]^\lw$ over nonlinear terms that contain products of two short wavelength pieces. Fortunately, the exact time interval and space volume over which the average $[\ldots]^\lw$ is taken is irrelevant if the parameter $\epsilon$ is sufficiently small.} This  {result} can be deduced from the formal equivalence between the
  local and global approaches to gyrokinetics established in reference
  \cite{Parra2015b}.}

The distribution function and electrostatic potential are written
using this decomposition,
\begin{eqnarray}
f_\spe& = f_\spe^\lw + f_\spe^\sw,\nonumber\\
\varphi& = \varphi^\lw
+ \varphi^\sw.
\end{eqnarray}

The separation of time and space scales suggests the following
assumptions on the distribution functions and electrostatic
potential. First, we assume that the relative sizes of the short and
long-wavelength components are
\begin{eqnarray} \label{orderings}
\frac{v_{t\spe}^3 f_\spe^\sw}{n_{e0}}\sim \frac{Z_\spe e
\varphi^\sw}{m_\spe v_{t\spe}^2} \sim \epsilon_s,
\nonumber\\[5pt]
\frac{v_{t\spe}^3 f_\spe^\lw}{n_{e0}}\sim
\frac{Z_\spe e \varphi^\lw}{m_\spe v_{t\spe}^2}
\sim 1,
\end{eqnarray}
where $n_{e0}$ is a characteristic value of the electron density. We
also need an ordering scheme for the space and time derivatives. The
natural assumptions are, for the long-wavelength components,
\begin{eqnarray}\label{orderings2}
\nabla_\boldr \ln f_\spe^\lw, \ \nabla_\boldr
\ln \varphi^\lw
\sim 1/L, \nonumber\\[5pt]
\partial_t \ln f_\spe^\lw ,\
\partial_t \ln \varphi^\lw \sim\epsilon_s^2 c_s/L,
\end{eqnarray}
and for the short-wavelength ones,
\begin{eqnarray}\label{orderings3}
\bun\cdot\nabla_\boldr \ln f_\spe^\sw,\ \bun\cdot\nabla_\boldr
\ln \varphi^\sw \sim 1/L,
\nonumber\\[5pt]
\nabla_{\boldr_\perp} \ln f_\spe^\sw,\ \nabla_{\boldr_\perp}
\ln \varphi^\sw \sim 1/\rho_s,
\nonumber\\[5pt]
\partial_t \ln f_\spe^\sw, \ \partial_t 
\ln \varphi^\sw \sim
c_s/L.
\end{eqnarray}
Here, $\perp$ stands for perpendicular to $\bB$ and $\bun:=B^{-1}\bB$
is the unit vector in the direction of $\bB$. Note the different sizes
of the perpendicular and the parallel gradients of $f_\spe^\sw$ and
$\varphi^\sw$.  {It is important to mention that the spatial dependence of $f_\spe^\sw$ and $\varphi^\sw$ has two different characteristic scales in the direction perpendicular to the magnetic field: the short length scale of the turbulent fluctuations, and the long length scale associated to the slow change of the turbulent average characteristics over the radius of the tokamak. These two different scales can be treated explicitly in the gyrokinetic equations \cite{Parra2015, Parra2015b}, but in this article we choose not to do it because we are deriving a global $\delta f$ gyrokinetic formulation.} 

Finally, we assume that the source term in equation \eq{eq:FPinitial}
only consists of the long-wavelength component, $S_\spe = S_\spe^\lw$,
and that it varies in the transport time scale $\tau_E$. For
consistency with the transport time scale $\tau_E$, we assume
\begin{equation}\label{eq:sizesource}
S_\spe \sim \epsilon_s^2\frac{n_{e0}c_s}{Lv_{t\spe}^3}.
\end{equation}

With the ordering detailed in this section, the gyromotion degree of
freedom can be averaged out order by order in $\epsilon_s$, and a
consistent set of non-linear equations for plasma microturbulence is
obtained~\cite{dorland00, dannert05, candy03, chen03, peeters09}.

%%%%%%%%%%%%%%%%%%%%%%%%%%%%%%%%%%%%%%%%%%%%%%%%%%%%%%%%%%%%%%%
\section{Radial flux of toroidal angular momentum}
\label{sec:radialFluxTorAngMom}
%%%%%%%%%%%%%%%%%%%%%%%%%%%%%%%%%%%%%%%%%%%%%%%%%%%%%%%%%%%%%%%

In this section we proceed as in references \cite{parra10a,
  Parra2015} to obtain a convenient expression for the radial
flux of toroidal angular momentum. In \cite{parra10a} a mass ratio
expansion in $\sqrt{m_e/m_i}\ll 1$ was performed in order to simplify
the presentation, that consequently only needed to treat kinetically
the ions (note that in \cite{parra11NF,Parra2015} the final
equations for both kinetic ions and kinetic electrons are
given). Here, we will not perform a mass ratio expansion.

We multiply the Fokker-Planck equation
\eq{eq:FPinitial} by $m_\spe \bv$ and integrate in velocity space,
obtaining
\begin{eqnarray}\label{eq:momentumEqspe}
\fl\partial_t (m_\spe {N}_\spe \bV_\spe) =
-\nabla_\boldr\cdot
\matP_\spe
-Z_\spe e{N}_\spe
\nabla_\boldr\varphi
+\frac{Z_\spe e}{c}
{N}_\spe\bV_\spe\times\bB
\nonumber\\[5pt]
\fl\hspace{1cm}
+\int m_\spe\bv\sum_{\spe'}C_{\spe \spe'}[f_\spe,f_{\spe'}]\dd^3 v +
\int m_\spe\bv S_\spe\dd^3 v
 .
\end{eqnarray}
The density, velocity and stress tensor of species $\spe$ are defined
by
\begin{equation}\label{eq:defdensity}
{N}_\spe (\boldr,t) := \int f_\spe(\boldr,\bv,t) \dd^3 v,
\end{equation}
\begin{equation}
\bV_\spe (\boldr,t) := \frac{1}{{N}_\spe}\int \bv f_\spe(\boldr,\bv,t) \dd^3 v,
\end{equation}
\begin{eqnarray}\label{eq:defPressureTensor}
\matP_\spe (\boldr,t)
:=
\int m_\spe \bv\bv f_\spe(\boldr,\bv,t)\dd^3 v.
\end{eqnarray}
 {Summing} \eq{eq:momentumEqspe} over species,
\begin{eqnarray}\label{eq:momentumEq}
\fl\partial_t \sum_\spe m_\spe {N}_\spe \bV_\spe =
-\nabla_\boldr\cdot
 \sum_\spe
\matP_\spe
+\frac{1}{c}
\bJ\times\bB + \sum_\spe\int m_\spe\bv S_\spe\dd^3 v,
\end{eqnarray}
where
\begin{equation}
\bJ (\boldr,t) := 
\sum_\spe Z_\spe e {N}_\spe\bV_\spe
\end{equation}
is the electric current density. To write \eq{eq:momentumEq} we have
employed the momentum conservation property of the collision operator
and the quasineutrality equation \eq{eq:quasineutrality}, that can be
expressed as $\sum_\spe Z_\spe e{N}_\spe = 0$.

We define the flux-surface average of a function
$G(\psi,\Theta,\zeta)$ by
\begin{equation}
\fl\hspace{0.5cm}\langle G \rangle_\psi :=
\frac{\int_0^{2\pi}\int_0^{2\pi}\sqrt{g}\,
 G(\psi,\Theta,\zeta) \dd\Theta\dd\zeta}
{\int_0^{2\pi}\int_0^{2\pi}
\sqrt{g}\,
 \dd\Theta\dd\zeta}\, ,
\end{equation}
with
\begin{equation}\label{eq:sqrtg}
\sqrt{g}:=\frac{1}{\nabla_\boldr\psi\cdot\left(
\nabla_\boldr\Theta\times\nabla_\boldr\zeta
\right)}
\end{equation}
the square root of the determinant of the metric tensor in
coordinates $\{\psi,\Theta,\zeta\}$. It will also be useful to define
the volume enclosed by the flux surface labeled by $\psi$,
\begin{equation}\label{eq:defvolume}
V(\psi) = 
 {\int_0^\psi\dd\psi^\prime\int_0^{2\pi}\dd\Theta\int_0^{2\pi}\dd\zeta
\, \sqrt{g} (\psi^\prime, \Theta)\, .}
\end{equation}
In order to find the equation for toroidal angular momentum transport we take
the scalar product of \eq{eq:momentumEq} with $R\zun =
R^2\nabla_\boldr\zeta$ and flux-surface average. We get
\begin{eqnarray}\label{eq:EvoltotalTorAngMom}
\fl\partial_t\left\langle
\sum_\spe  m_\spe {N}_\spe R\zun\cdot \bV_\spe
\right\rangle_\psi
&=
-\frac{1}{V'}\partial_\psi 
(V'\Pi)
+ T_\zeta
,
\end{eqnarray}
where the prime stands for differentiation with respect to $\psi$,
the external torque is
\begin{equation}
T_\zeta  {:=} \left\langle \sum_\spe R \int m_\spe\bv\cdot\zun
 S_\spe\dd^3 v\right\rangle_\psi,
\end{equation}
and the radial flux of toroidal angular momentum is defined by
\begin{equation}\label{eq:Pi}
\Pi := \sum_\spe\Pi_\spe,
\end{equation}
with
\begin{equation}
\Pi_\spe
:=
\left\langle
R \zun\cdot
\matP_\spe\cdot\nabla_\boldr\psi
\right\rangle_\psi.
\end{equation}
We have used that $\nabla_\boldr(R\zun)$ is antisymmetric and that
$\left\langle R \zun\cdot(\bJ\times\bB) \right\rangle_\psi =
\left\langle \bJ\cdot\nabla_\boldr\psi \right\rangle_\psi $. The latter
has to be zero, as can be proven by noting that $\nabla_\boldr\cdot\bJ =
0$, and that $\bJ\cdot\nabla_\boldr\psi$ must vanish at $\psi = 0$.

In the low flow ordering, the magnitude of the ion flow is $V_i\sim
\epsilon_s c_s$. Recalling the definition of the transport time scale
given in Section \ref{sec:orderingsandassumptions}, $\tau_E =
\epsilon_s^{-2}L/c_s$, we can estimate the size of the first term in
\eq{eq:EvoltotalTorAngMom}. Namely,
\begin{eqnarray}
\fl\partial_t\left\langle
\sum_\spe  m_\spe {N}_\spe R\zun\cdot \bV_\spe
\right\rangle_\psi
\sim \epsilon_s^3 \frac{R n_{e0} T_0}{L}.
\end{eqnarray}
If the three terms in equation \eq{eq:EvoltotalTorAngMom} are to be
comparable, we deduce
\begin{equation}
\Pi\sim \epsilon_s^3 n_{e0}{T_0} R B_0 L
\end{equation}
and
\begin{equation}\label{eq:sizeTzeta}
T_\zeta\sim \epsilon_s^3 \frac{R n_{e0} T_0}{L}.
\end{equation}

Observe that we are assuming that the momentum injection is smaller in
$\epsilon_s$ than, for example, the particle input (see
\eq{eq:sizesource}). If $T_\zeta$ were much larger than the estimate
\eq{eq:sizeTzeta}, it would dominate the toroidal rotation. Then,
$V_i$ would approach sonic values and the plasma would enter the
so-called high-flow regime. As pointed out in the Introduction, it is
unlikely that momentum input larger than the estimate
\eq{eq:sizeTzeta} be available in the core of large tokamaks, and
therefore the only possible rotation will be of intrinsic
nature. However, this is precisely the difficult case from the
theoretical and computational perspective: in the low-flow regime, one
needs to calculate $\Pi$ to $O(\epsilon_s^3)$.

In references \cite{parra10a, Parra2015}, a procedure was
devised to write $\Pi$ in such a way that `only' $O(\epsilon_s^2)$
pieces of the distribution functions and electrostatic potential are
needed to evaluate it. We start by multiplying \eq{eq:FPinitial} by
$m_\spe\bv\bv$ and integrating over velocities,
\begin{eqnarray}\label{eq:relationP}
\fl
\Omega_\spe
\left(
\matP_\spe\times\bun - \bun\times\matP_\spe 
\right)
=
\nonumber\\[5pt]
\fl
\hspace{1cm}
\partial_t\matP_\spe
+
\nabla_\boldr
\cdot
\left(
m_\spe\int \bv\bv\bv f_\spe\dd^3 v
\right)
\nonumber\\[5pt]
\fl
\hspace{1cm}+
Z_\spe e {N}_\spe\left(
\nabla_\boldr\varphi \bV_\spe + \bV_\spe\nabla_\boldr\varphi
\right)
\nonumber\\[5pt]
\fl
\hspace{1cm}
-
\int
m_\spe\bv\bv
\sum_{\spe'}C_{\spe \spe'}[f_\spe,f_{\spe'}]
\dd^3 v
\nonumber\\[5pt]
\fl
\hspace{1cm}
-
\int
m_\spe\bv\bv
S_\spe
\dd^3 v
.
\end{eqnarray}

We are only interested in the long-wavelength component of this
equation. We find the expression for $\Pi_\spe^\lw$ by taking the
double-dot product of \eq{eq:relationP} with $R^2\zun\zun/2$,
flux-surface averaging, and extracting the long-wavelength component,
\begin{eqnarray}\label{eq:Pidimpreliminary}
\fl
\Pi_\spe^\lw
=
\frac{m_\spe c}{2 Z_\spe e}\partial_t
\left\langle
R^2 \hat{\zetabf}\cdot\matP_\spe^\lw\cdot\hat{\zetabf}
\right\rangle_\psi
\nonumber\\[5pt]
\fl
\hspace{1cm}
+
\frac{m_\spe^2c}{2Z_\spe e}
\frac{1}{V'}\partial_\psi
\left\langle
V'
\int
f_\spe^\lw(\bv\cdot\nabla_\boldr\psi)R^2(\bv\cdot\zun)^2
\dd^3v
\right\rangle_\psi
\nonumber\\[5pt]
\fl
\hspace{1cm}
+
\left\langle
c\partial_\zeta
\varphi
R {N}_\spe m_\spe (\bV_\spe\cdot\hat{\zetabf})
\right\rangle_\psi^\lw
\nonumber\\[5pt]
\fl\hspace{0.5cm}
-
\frac{m_\spe^2 c}{2 Z_\spe e}
\left\langle
\int 
\sum_{\spe'}C_{\spe\spe'}^\lw
R^2 (\bv\cdot\hat{\zetabf})^2
\dd^3 v
\right\rangle_\psi
\nonumber\\[5pt]
\fl\hspace{0.5cm}
-
\frac{m_\spe^2 c}{2 Z_\spe e}
\left\langle
\int 
S_\spe
R^2 (\bv\cdot\hat{\zetabf})^2
\dd^3 v
\right\rangle_\psi
.
\end{eqnarray}

In references \cite{parra10a, Parra2015} it is shown that, up to
terms that are small by a factor of $\epsilon_s$, the right-hand side
of \eq{eq:Pidimpreliminary} can be rewritten as
\begin{eqnarray}\label{eq:Pidim}
\fl
\Pi_\spe^\lw
=
\left\langle
c\partial_\zeta
\varphi
R {N}_\spe m_\spe (\bV_\spe\cdot\hat{\zetabf})
\right\rangle_\psi^\lw
\nonumber\\[5pt]
\fl\hspace{0.5cm}
+
\frac{m_\spe c^2}{2 Z_\spe e}
\frac{1}{V'}\partial_\psi
\left\langle
V'
\partial_\zeta\varphi R^2(\hat{\zetabf}\cdot\matP_\spe\cdot\hat{\zetabf})
\right\rangle_\psi^\lw
+
\frac{m_\spe c}{2 Z_\spe e}\langle R^2 
\rangle_\psi\partial_t p_\spe
\nonumber\\[5pt]
\fl\hspace{0.5cm}
-
\frac{m_\spe^2 c}{2 Z_\spe e}
\left\langle
\int 
\sum_{\spe'}C_{\spe\spe'}^\lw
R^2 (\bv\cdot\hat{\zetabf})^2
\dd^3 v
\right\rangle_\psi
\nonumber\\[5pt]
\fl\hspace{0.5cm}
-
\frac{m_\spe^3 c^2}{6 Z_\spe^2 e^2}
\frac{1}{V'}\partial_\psi
\left\langle
V'
\int 
\sum_{\spe'}C_{\spe\spe'}^\lw
R^3 (\bv\cdot\hat{\zetabf})^3
\dd^3 v
\right\rangle_\psi
\nonumber\\[5pt]
\fl\hspace{0.5cm}
-
\frac{m_\spe^2 c}{2 Z_\spe e}
\left\langle
\int 
S_\spe
R^2 (\bv\cdot\hat{\zetabf})^2
\dd^3 v
\right\rangle_\psi
,
\end{eqnarray}
where $p_\spe  {= n_\sigma T}$ is the pressure of species  {$\spe$,} given by the
  $O(n_{e0}{T_0})$ term of the right side of
  \eq{eq:defPressureTensor}. In Section
\ref{sec:GyrokineticExpansion} we explicitly show that the right-hand
side of \eq{eq:Pidim} can be recast in terms of the fields of the
theory to $O(\epsilon_s^2)$ by rewriting it in such a way that the
solutions of the gyrokinetic equations (derived in Sections
\ref{sec:SWequations} and \ref{sec:LongWave}) can be directly plugged
into it.

%%%%%%%%%%%%%%%%%%%%%%%%%%%%%%%%%%%%%%%%%%%%%%%%%%%%%%%%%%%%%%%%%%%%%%%
\section{Gyrokinetic expansion}
\label{sec:GyrokineticExpansion}
%%%%%%%%%%%%%%%%%%%%%%%%%%%%%%%%%%%%%%%%%%%%%%%%%%%%%%%%%%%%%%%%%%%%%%%

The objective is to write the right side of \eq{eq:Pidim} in terms of
the pieces of the distribution functions and electrostatic potential
that are obtained by solving the gyrokinetic equations. This is done
in subsection \ref{sec:radialfluxtoroidalmomentum}. Before doing this,
in subsections \ref{sec:dimensionlessvariables} and
\ref{sec:gyrokinCoorTransf} we briefly recall how the gyrokinetic
expansion is defined and give the results of references
\cite{CalvoParra2012} and \cite{ParraCalvo2011} that are needed in
this article.

%%%%%%%%%%%%%%%%%%%%%%%%%%%%%%%%%%%%%%%%%%%%%%%%%%%%%%%%%%%%%%%
\subsection{Dimensionless variables}
\label{sec:dimensionlessvariables}
%%%%%%%%%%%%%%%%%%%%%%%%%%%%%%%%%%%%%%%%%%%%%%%%%%%%%%%%%%%%%%%

The gyrokinetic expansion is more clearly understood by writing the
equations in dimensionless variables. For time, space, magnetic field,
electrostatic potential, particle density  {and temperature}, we employ
\begin{eqnarray} \label{norm_spindep}
\underline{t} = \frac{c_s t}{L}, \ \underline{\boldr} =
\frac{\boldr}{L}, \ \underline{\bB} = \frac{\bB}{B_0},
\nonumber\\[5pt]
\underline{\varphi} = \frac{e \varphi}{\epsilon_s T_{0}}, \
\underline{n_{\sigma}}=\frac{n_\sigma}{n_{e0}}, \
\underline{T_{\sigma}}=\frac{T_\spe}{T_{0}}.
\end{eqnarray}
For  {velocities, distribution functions and sources,} we use
\begin{equation} \label{norm_spdep}
\underline{\bv_\spe} = \frac{\bv_\spe}{v_{t\spe}}, \
\underline{f_\spe} = \frac{v_{t\spe}^3}{n_{e0}} f_\spe, \
 {\underline{S_\spe} = \frac{L}{\epsilon_s^2 c_s}\frac{v_{t\spe}^3}{n_{e0}}S_\spe.}
\end{equation}

In dimensionless variables, the Fokker-Planck
equation~\eq{eq:FPinitial} becomes
\begin{eqnarray}\label{eq:FPnon-dim}
\fl
\partial_{\underline{t}}\, \underline{f_\spe} +
\tau_\spe
\underline{\bv}\cdot
\nabla_{\underline{\boldr}}\, \underline{f_\spe}
+\tau_\spe
\left(
-Z_\spe\epsilon_s\nabla_{\underline{\boldr}}\underline{\varphi}
+
\frac{1}{\epsilon_\spe}\underline{\bv}\times\underline{\bB}
\right)
\cdot
\nabla_{\underline{\bv}}\,\underline{f_\spe}
\nonumber\\[5pt]
\fl\hspace{1cm}
 = \tau_\spe
\sum_{\spe'}\underline{C_{\spe \spe'}}
[\underline{f_\spe},\underline{f_{\spe'}}]
(\underline{\boldr},\underline{\bv}) + \epsilon_s^2\underline{S_\spe}.
\end{eqnarray}
The normalized collision operator reads, explicitly,
\begin{eqnarray}\label{eq:collisionoperatornondim}
\fl\underline{C_{\spe \spe'}}
[\underline{f_\spe},\underline{f_{\spe'}}]
(\underline{\boldr},\underline{\bv}) =\nonumber\\[5pt]
\fl\hspace{1cm}\underline{\gamma_{\spe\spe'}} \nabla_{\underline{\bv}}\cdot
\int \matW\left(\tau_\spe\underline{\bv} -
\tau_{\spe^\prime}\underline{\bv'}\right) \cdot
\Bigg( \tau_\spe\underline{f_{\spe'}}(\underline{\boldr},\underline{\bv'},
\underline{t})
\nabla_{\underline{\bv}}\,
\underline{f_\spe}(\underline{\boldr},\underline{\bv},\underline{t})
\nonumber\\[5pt]
\fl\hspace{1cm}
-
\tau_{\spe'}
\underline{f_{\spe}}(\underline{\boldr},\underline{\bv},\underline{t})
\nabla_{\underline{\bv'}} \,\underline{f_{\spe'}}
(\underline{\boldr},\underline{\bv'},\underline{t}) \Bigg)
\dd^3\underline{v'}.\nonumber\\
\end{eqnarray}
Here,
\begin{equation}
\underline{\gamma_{\spe\spe'}}:= \frac{2 \pi
Z_\spe^2 Z_{\spe'}^2 n_{e0} e^4 L}{{T_0}^2}\ln\Lambda.
\end{equation}
is the collisionality (up to a factor of order unity with respect
standard definitions). In this paper we assume
$\underline{\gamma_{\spe\spe'}}\sim 1$.

The quasineutrality equation \eq{eq:quasineutrality} is recast as
\begin{eqnarray}\label{eq:Quasineutralitynondim}
 \sum_\spe Z_\spe
\int \underline{f_\spe}(\underline{\boldr}, \underline{\bv},
\underline{t}) \dd^3 \underline{v}=0.
\end{eqnarray} 

It is useful to list the ordering assumptions \eq{orderings},
\eq{orderings2} and \eq{orderings3} in terms of dimensionless
variables. The short-wavelength components of the electrostatic
potential and the distribution functions satisfy  {
\begin{eqnarray}\label{eq:ordering_sw_dimensionless}
\underline{\varphi}^\sw(\underline{\boldr}, \underline{t})\sim 1,
\nonumber\\[5pt]
\underline{f_\spe^\sw} (\underline{\boldr},\underline{\bv},
\underline{t})
\sim \epsilon_s,\nonumber\\[5pt]
{\bun}(\underline{\boldr}) \cdot \nabla_{\underline{\boldr}}\,
\underline{\varphi}^\sw (\underline{\boldr}, \underline{t}) \sim 1,
\nonumber\\[5pt]
{\bun}(\underline{\boldr}) \cdot \nabla_{\underline{\boldr}}\,
\underline{f_\spe^\sw} (\underline{\boldr},\underline{\bv},
\underline{t}) \sim \epsilon_s,
\nonumber\\[5pt]
 \partial_{\underline{t}} \underline{\varphi}^\sw (\underline{\boldr}, \underline{t}) \sim 1,
\nonumber\\[5pt] \partial_{\underline{t}} \underline{f_\spe^\sw} (\underline{\boldr},\underline{\bv},
\underline{t}) \sim \epsilon_s.
\end{eqnarray}}

 {The} ordering for the long-wavelength components is
\begin{eqnarray}\label{eq:ordering_lw_dimensionless}
  \underline{\varphi}^\lw(\underline{\boldr},
  \underline{t})\sim 1/\epsilon_s,\nonumber\\[5pt]
  \underline{f_\spe^\lw}(\underline{\boldr},\underline{\bv},
  \underline{t})\sim 1,\nonumber\\[5pt]
  \nabla_{\underline{\boldr}}\, \underline{\varphi}^\lw
  (\underline{\boldr},
  \underline{t})  \sim 1/\epsilon_s,\nonumber\\[5pt]
  \nabla_{\underline{\boldr}}\, \underline{f_\spe^\lw}
  (\underline{\boldr},
  \underline{t})  \sim 1,
  \nonumber\\[5pt] \partial_{\underline{t}} \underline{\varphi}^\lw (\underline{\boldr}, \underline{t}) \sim \epsilon_s,
  \nonumber\\[5pt] \partial_{\underline{t}} \underline{f_\spe^\lw}(\underline{\boldr},\underline{\bv},
  \underline{t}) \sim \epsilon_s^2.
\end{eqnarray}
Observe that the normalization of $\varphi$ has an extra
$\epsilon_s^{-1}$ that makes $\underline{\varphi}^\lw\sim
\epsilon_s^{-1}$ and $\underline{\varphi}^\sw\sim 1$, whereas
$\underline{f_\spe^\lw}\sim 1$ and $\underline{f_\spe^\sw}\sim
\epsilon_s$. For the expansions  {$\epsilon_s \ll 1$} of both components of the normalized
electrostatic potential, we write
\begin{equation}\label{eq:potentiallw1}
\underline{\varphi}^\lw(\underline{\boldr},
\underline{t}) :=
\frac{1}{\epsilon_s}\underline{\varphi_0}(\underline{\boldr},
\underline{t}) +
\underline{\varphi_1}^\lw(\underline{\boldr},
\underline{t}) + \epsilon_s
\underline{\varphi_2}^\lw(\underline{\boldr},
\underline{t}) +
O(\epsilon_s^2)
\end{equation}
and
\begin{equation}\label{eq:potentialsw1}
\underline{\varphi}^\sw(\underline{\boldr},
\underline{t}) :=
\underline{\varphi_1}^\sw(\underline{\boldr},
\underline{t}) + \epsilon_s
\underline{\varphi_2}^\sw(\underline{\boldr},
\underline{t}) +
O(\epsilon_s^2).
\end{equation}

We do not underline variables from here on, but we assume that we are
employing the non-dimensional ones.

%%%%%%%%%%%%%%%%%%%%%%%%%%%%%%%%%%%%%%%%%%%%%%%%%%%%%%%%%%%%%%%
\subsection{Gyrokinetic coordinate transformation}
\label{sec:gyrokinCoorTransf}
%%%%%%%%%%%%%%%%%%%%%%%%%%%%%%%%%%%%%%%%%%%%%%%%%%%%%%%%%%%%%%%

The aim of gyrokinetic theory is to average
equations \eq{eq:FPnon-dim} and \eq{eq:Quasineutralitynondim} over
time scales of the order of the ion gyrofrequency, assuming that the
orderings of Section \ref{sec:orderingsandassumptions} hold. The
typical approaches to the problem try to find a coordinate
transformation in phase space such that in the new coordinates, called
gyrokinetic coordinates, the fast degree of freedom is
decoupled. Iterative methods work directly on the equations of
motion~\cite{frieman82, lee83, wwlee83, bernstein85, parra08}, whereas
Hamiltonian and Lagrangian methods employ techniques of analytical
mechanics to obtain the coordinate transformation~\cite{dubin83,
  hahm88, brizard07,ParraCalvo2011}.

It has been proven in \cite{parra10a} that the intrinsic ambipolarity
of the tokamak implies that the gyrokinetic equations have to be
computed to $O(\epsilon_s^2)$ if one wants to find the long-wavelength
radial electric field. The same accuracy is needed to compute the
intrinsic rotation profile in the low-flow ordering. The calculation
of the gyrokinetic system of equations to second order is given in
reference \cite{ParraCalvo2011}, in the phase-space Lagrangian
formalism, for general magnetic geometry.  {The correctness of
  the approach and the results in \cite{ParraCalvo2011} has recently
  been confirmed by independent calculations of the most involved
  piece of the Lagrangian~\cite{Burby2013,Parra2014}. In simplified
  geometries, the results of \cite{ParraCalvo2011} reduce to those of
  seminal references like \cite{dubin83}.}

 {However,} reaching expressions sufficiently explicit to be
implemented in a code still requires some work. In reference
\cite{CalvoParra2012} this was done for tokamak geometry but only for
the equations that give the long-wavelength components of the
fields. Here, we introduce some notation and refresh some results from
\cite{CalvoParra2012} and \cite{ParraCalvo2011} that will be needed in
the following sections, where the equations determining the
short-wavelength components of the distribution functions and
electrostatic potential are derived.

We denote by $(\bR,u,\mu,\theta)$ the gyrokinetic coordinates, where
$\bR$ is the position of the gyrocenter, $u$ is the parallel velocity,
$\mu$ is the magnetic moment and $\theta$ is the gyrophase. The
euclidean phase-space coordinates are denoted by
$\bX\equiv\{\boldr,\bv\}$ and the transformation between both sets of
coordinates by $\cT_\spe$,
\begin{equation}
(\boldr,\bv) = \cT_\spe(\bR,u,\mu,\theta,t).
\end{equation}
In practice, the transformation is computed as a power series in
$\epsilon_\spe$, where the lowest order terms of $\cT_\spe^{-1}$ are
given by
\begin{eqnarray}
\bR = \boldr
 -\epsilon_\spe\frac{1}{B(\boldr)}\bun(\boldr)\times\bv
+ O(\epsilon_\spe^2)
,\nonumber\\[5pt]
u = \bv\cdot\bun(\boldr) + O(\epsilon_\spe)
,\nonumber\\[5pt]
\mu = \frac{1}{2 B(\boldr)}
\left(\bv - \bv\cdot\bun(\boldr)\bun(\boldr)\right)^2
+ O(\epsilon_\spe)
,\nonumber\\[5pt]
\theta = \arctan
\left(\frac{\bv\cdot\eun_2(\boldr)}{\bv\cdot\eun_1(\boldr)}
\right)
+ O(\epsilon_\spe).
\end{eqnarray}
Here, the unit vectors $\eun_1 (\boldr)$ and $\eun_2 (\boldr)$ are
orthogonal to each other and to $\bun (\boldr)$, and satisfy $\eun_1
(\boldr) \times \eun_2 (\boldr) = \bun (\boldr)$ at every location
$\boldr$. 

It will be useful to write some expressions in terms of $\cT^*_\spe$,
the pull-back transformation induced by $\cT_\spe$. Given a function
$g(\bX, t)$, $\cT_\spe^* g (\bZ,t)$ is
\begin{equation}
\cT_\spe^* g (\bZ,t) = g(\cT_\spe(\bZ,t), t).
\end{equation}

For the expansion of the transformation in powers of $\epsilon_\spe$ we employ the notation
\begin{eqnarray}
  \cT_\spe =  \cT_{\spe,0} + \epsilon_\spe \cT_{\spe,1} 
+ \epsilon_\spe^2 \cT_{\spe,2} + O(\epsilon_\spe^3),\nonumber\\[5pt]
 \cT_\spe^{-1} =  \cT^{-1}_{\spe,0} + \epsilon_\spe \cT^{-1}_{\spe,1} 
+ \epsilon_{\spe}^2 \cT^{-1}_{\spe,2} + O(\epsilon_\spe^3).
\end{eqnarray}
We will make an extensive use of the zeroth-order transformation. The
expression for $(\bR,u,\mu,\theta) = \cT_{\spe,0}^{-1}(\boldr,\bv)$ is
\begin{eqnarray}\label{eq:TinvZerothOrder}
\bR = \boldr
,\nonumber\\[5pt]
u = \bv\cdot\bun(\boldr)
,\nonumber\\[5pt]
\mu = \frac{1}{2 B(\boldr)}
\left(\bv - \bv\cdot\bun(\boldr)\bun(\boldr)\right)^2
,\nonumber\\[5pt]
\theta = \arctan
\left(\frac{\bv\cdot\eun_2(\boldr)}{\bv\cdot\eun_1(\boldr)}
\right)
.
\end{eqnarray}
Higher-order terms of the change of coordinates will be computed or
taken from \cite{CalvoParra2012} when needed. Given a function
$G(\bR,u,\mu,\theta)$, the following obvious identity will also be
useful,
\begin{eqnarray}
\int \cT_{\spe,0}^{-1*}G(\boldr,\bv)\dd^3 v =
\int B(\boldr) G(\boldr,u,\mu,\theta)\dd u \dd\mu\dd\theta,
\end{eqnarray}
where we have used that the Jacobian of $\cT_{\spe,0}$ is
$B(\boldr)$.

Applying the gyrokinetic transformation to the the Fokker-Planck
equation \eq{eq:FPnon-dim}, we get
\begin{eqnarray}\label{eq:FokkerPlancknondimgyro}
\fl\partial_{t} F_\spe
 +
\tau_\spe\dot\bR\cdot\nabla_\bR F_\spe
+
\tau_\spe\dot u\partial_u F_\spe
+ \tau_\spe\dot\theta\partial_\theta F_\spe
\nonumber\\[5pt]
\fl\hspace{1cm}
 =
\tau_\spe \sum_{\spe'}\cT^{*}_{\spe} C_{\spe
\spe'} [\cT^{-1*}_{\spe} F_\spe,\cT^{-1*}_{\spe'}F_{\spe'}]
(\bZ, t) + \epsilon_s^2\cT_\spe^* S_\spe,
\end{eqnarray}
where $ F_\spe:=\cT_\spe^* f_\spe $, $\cT^{-1*}_{\spe}$ is the
pull-back transformation that corresponds to $\cT_\spe^{-1}$,
i.e. $\cT_\spe^{-1*} F_\spe (\bX,t) = F_\spe(\cT_\spe^{-1}(\bX,t),t)$,
and the particle equations of motion, $\dot\bR$, $\dot u$, and
$\dot\theta$ (note that $\dot\mu = 0$) are given in
\ref{sec:eqsofmotion} to the necessary order.

Since it will be also useful in this paper, we recall that in
\cite{ParraCalvo2011} the gyrokinetic transformation $\cT_\spe$ is
written as the composition of two transformations, $ \cT_\spe =
\cT_{NP,\spe} \cT_{P,\spe}$. First, we have the {\it non-perturbative
  transformation} $(\boldr, \bv) = \cT_{NP, \spe} (\bZ_g) \equiv
\cT_{NP,\spe} (\bR_g, v_{||g}, \mu_g, \theta_g)$, defined by
\begin{eqnarray}\label{eq:changeNonPert}
\boldr &=& \bR_g  + \epsilon_\spe \rhobf ( \bR_g, \mu_g, \theta_g),
\nonumber\\[5pt]
\bv &=& v_{||g} \bun(\bR_g) + \rhobf ( \bR_g, \mu_g, \theta_g)
\times \bB (\bR_g),
\end{eqnarray}
with the gyroradius vector defined as
\begin{equation}\label{eq:defrho}
\rhobf ( \bR_g, \mu_g, \theta_g ) = - \sqrt{\frac{2
\mu_g}{B(\bR_g)}} \left [ \sin \theta_g \eun_1 (\bR_g) - \cos
\theta_g \eun_2 (\bR_g) \right ].
\end{equation}
Second, the {\it perturbative transformation}
\begin{equation}\label{eq:defPertTransf}
(\bR_g, v_{||g}, \mu_g, \theta_g) = \cT_{P, \spe} (\bR, u, \mu, \theta, t),
\end{equation}
that is written as a power series in $\epsilon_\spe$,
\begin{eqnarray}\label{newvar}
\bR_g &= \bR + \epsilon_\spe^{2} {\bR}_{\spe,2} + O(\epsilon_\spe^3),
\nonumber\\[5pt]
v_{||g} &= u + \epsilon_\spe {u}_{\spe,1}, + O(\epsilon_\spe^2),
\nonumber\\[5pt]
\mu_g &=
\mu + \epsilon_\spe {\mu}_{\spe,1}  + O(\epsilon_\spe^2),
\nonumber\\[5pt]
\theta_g &= \theta
+ \epsilon_\spe {\theta}_{\spe,1}  + O(\epsilon_\spe^2).
\end{eqnarray}
The corrections ${\bR}_{\spe,2}$, ${u}_{\spe,1}$,
${\mu}_{\spe,1}$ and ${\theta}_{\spe,1}$ are given in
\ref{sec:lowestordertermsTP}.

In gyrokinetic variables the quasineutrality equation
\eq{eq:Quasineutralitynondim} reads
\begin{eqnarray}\label{eq:gyroQuasineutrality}
\fl\sum_\spe Z_\spe  \int |\det\left(J_{\spe}\right)|
F_\spe
\delta\Big(\pi^{\boldr}\Big(\cT_{\spe}(\bZ,
t)\Big)-\boldr\Big)\dd ^6Z = 0,
\end{eqnarray}
where $\pi^\boldr(\boldr,\bv):=\boldr$, and the Jacobian of $\cT_\spe$
to $O(\epsilon_\spe ^2)$ is
\begin{eqnarray}\label{eq:Jacobian}
|\det(J_{\spe} )| = B_{||,\spe}^*,
\end{eqnarray}
with $B_{||,\spe}^*$ defined in \eq{Bstar_par}.

Finally, we recall that the gyrokinetic equations are written naturally
in terms of a function $\phi_\spe$ defined as
\begin{equation}
\phi_\spe(\bR,\mu,\theta,t) :=
\varphi(\bR+\epsilon_\spe\rhobf(\bR,\mu,\theta),t).
\end{equation}
It is useful to introduce its gyrophase-independent piece
$\langle\phi_\spe\rangle$ and its gyrophase-dependent one
$\tilde\phi_\spe$,
\begin{equation}
\tilde\phi_\spe(\bR,\mu,\theta,t) :=
\phi_\spe(\bR,\mu,\theta,t)
 - \langle\phi_\spe\rangle(\bR,\mu,t).
\end{equation}
The gyroaverage of a function $G(\bR,u,\mu,\theta,t)$ is defined by
\begin{equation}
\langle
G
\rangle(\bR,u,\mu,t) :=
\frac{1}{2\pi}\int_0^{2\pi}G(\bR,u,\mu,\theta,t)
\dd\theta.
\end{equation}

The ordering assumptions on $\varphi$,
 {equations~\eq{eq:ordering_sw_dimensionless},} imply that
\begin{eqnarray}\label{eq:notationphisw}
  \langle\phi^\sw_\spe\rangle = \langle\phi^\sw_{\spe 1}\rangle 
  + \epsilon_s\langle\phi^\sw_{\spe 2}\rangle + O(\epsilon_s^2),\nonumber\\
  \tilde\phi^\sw_\spe = \tilde\phi^\sw_{\spe 1} 
+ \epsilon_s\tilde\phi^\sw_{\spe 2} + O(\epsilon_s^2).
\end{eqnarray}
A Taylor expansion of $\phi^\lw_\spe$ around $\boldr = \bR$ is
allowed. Employing the convention $\mathbf u\mathbf v :\matM = \mathbf
v \cdot\matM \cdot \mathbf u$ for an arbitrary matrix $\matM$ and
vectors $\mathbf u$ and $\mathbf v$, we write
\begin{eqnarray}\label{eq:potentiallw2}
\fl\langle\phi_\spe^\lw\rangle(\bR,\mu,t) =
\frac{1}{\epsilon_s}\varphi_0(\bR,t)
+\varphi_1^\lw(\bR,t)
\nonumber\\[5pt]
\fl
\hspace{0.5cm}
+\epsilon_s \Bigg( \frac{\mu}{2 Z_\spe^2\tau_\spe^2 B(\bR)}
 (\matI-\bun(\bR)\bun(\bR)):\nabla_{\bR}\nabla_{\bR}
\varphi_0(\bR,t)
\nonumber\\[5pt]
\fl
\hspace{0.5cm}
+\varphi_2^\lw(\bR,t) \Bigg)
 +O(\epsilon_s^2)
\end{eqnarray}
and
\begin{equation}\label{eq:potentiallw3}
\fl\tilde\phi_\spe^\lw(\bR,\mu,\theta,t) =
\frac{1}{Z_\spe\tau_\spe} \rhobf(\bR,\mu,\theta)
\cdot\nabla_{\bR}\varphi_0(\bR,t) + O(\epsilon_s),
\end{equation}
giving $\tilde\phi^\lw_\spe = O(1)$.

%%%%%%%%%%%%%%%%%%%%%%%%%%%%%%%%%%%%%%%%%%%%%%%%%%%%%%%%%%%%%%%
\subsection{Gyrokinetic expansion of the radial flux of toroidal
  angular momentum}
\label{sec:radialfluxtoroidalmomentum}
%%%%%%%%%%%%%%%%%%%%%%%%%%%%%%%%%%%%%%%%%%%%%%%%%%%%%%%%%%%%%%%

We start by writing the right-hand side of equation \eq{eq:Pidim} in
terms of dimensionless variables. We use the normalizations
\begin{eqnarray}\label{eq:normPiVP}
\Pi = 
n_{e0} T_0 B_0 L^2\underline{\Pi}
\nonumber\\[5pt]
\Pi_\spe = 
n_{e0} T_0 B_0 L^2\underline{\Pi_\spe}
\nonumber\\[5pt]
 {N_\spe \bV_\spe = \epsilon_s n_{e0}
c_s\underline{N_\spe \bV_\spe}}
\nonumber\\[5pt]
\matP_\spe = 
n_{e0} T_0\underline{\matP_\spe}.
\end{eqnarray}
Thus, in dimensionless variables (we do not underline them anymore),
\begin{eqnarray}
 {N_\spe} \bV_\spe (\boldr,t)
:=
\frac{\tau_\spe}{\epsilon_s}
\int \bv f_\spe(\boldr,\bv,t)\dd^3 v
\end{eqnarray}
and
\begin{eqnarray}
\matP_\spe (\boldr,t)
:=
\int \bv\bv f_\spe(\boldr,\bv,t)\dd^3 v.
\end{eqnarray}
 {Also,} we use the conventions
\begin{eqnarray}
 {N_\spe \bV_\spe = (N_\spe \bV_\spe)_1 + \epsilon_\spe (N_\spe \bV_\spe)_2 }
+ O(\epsilon_\spe^2)
\end{eqnarray}
and
\begin{eqnarray}
  \matP_\spe \ = \ \matP_{\spe 0}  + \epsilon_\spe \matP_{\spe 1} + \epsilon_\spe^2 \matP_{\spe 2} 
+ O(\epsilon_\spe^3).
\end{eqnarray}

In these variables, equation \eq{eq:Pidim} reads
\begin{eqnarray}\label{eq:Pinondim}
\fl
\Pi_\spe^\lw
=
\frac{\epsilon_s^2}{\tau_\spe^2}
\left\langle
\partial_{\zeta/\epsilon_s}
\varphi
R {N}_\spe (\bV_\spe\cdot\hat{\zetabf})
\right\rangle_\psi
^\lw
\nonumber\\[5pt]
\fl\hspace{0.5cm}
+
\frac{\epsilon_s^2 }{2Z_\spe\tau_\spe^2}
\frac{1}{V'}\partial_\psi
\left\langle
V'
\partial_{\zeta/\epsilon_s}
\varphi
R^2 (\hat{\zetabf}\cdot\matP_\spe\cdot\hat{\zetabf})
\right\rangle_\psi
^\lw
\nonumber\\[5pt]
\fl\hspace{0.5cm}
+
\frac{\epsilon_s^3}{2Z_\spe\tau_\spe^2}\langle R^2 \rangle_\psi
\partial_{\epsilon_s^2 t} p_\spe
-
\frac{\epsilon_s}{2Z_\spe\tau_\spe}
\left\langle
\int 
\sum_{\spe'}C_{\spe\spe'}^\lw
R^2 (\bv\cdot\hat{\zetabf})^2
\dd^3 v
\right\rangle_\psi
\nonumber\\[5pt]
\fl\hspace{0.5cm}
-
\frac{\epsilon_s^2 }{6Z_\spe^2\tau_\spe^2}
\frac{1}{V'}
\partial_\psi
\left\langle
V'
\int 
\sum_{\spe'}C_{\spe\spe'}^\lw
R^3 (\bv\cdot\hat{\zetabf})^3
\dd^3 v
\right\rangle_\psi
\nonumber\\[5pt]
\fl\hspace{0.5cm}
-
\frac{\epsilon_s^3}{2Z_\spe\tau_\spe^2}
\left\langle
\int 
S_\spe
R^2 (\bv\cdot\hat{\zetabf})^2
\dd^3 v
\right\rangle_\psi,
\end{eqnarray}
where we have omitted the arguments of the collision operators to
ease the notation slightly.

We turn to write the right-hand side of \eq{eq:Pinondim} in terms of
the solutions obtained from the gyrokinetic Fokker-Planck
and quasineutrality equations that are given in Sections
\ref{sec:FPandQuasineutSWfirstOrder},
\ref{sec:shortWaveFokkerPlanckEq}, \ref{sec:shortWavePoissonEq} and
\ref{sec:LongWave}. From the Fokker-Planck equation one obtains, order
by order, $F_{\spe 0}$, $F_{\spe 1}$ and $F_{\spe 2}$, where our
convention is
\begin{eqnarray}
\fl
F_\spe(\bR,u,\mu,\theta) = 
F_{\spe 0}(\bR,u,\mu) +
\epsilon_\spe F_{\spe 1}(\bR,u,\mu)
\nonumber\\[5pt]
\fl\hspace{1cm}+
\epsilon_\spe^2 F_{\spe 2}(\bR,u,\mu,\theta) + O(\epsilon_\spe^3).
\end{eqnarray}
In reference \cite{CalvoParra2012}, it is shown that $F_{\spe 1}$ does
not depend on the gyrophase and that $F_{\spe 0}$ is a Maxwellian with
zero flow, whose density and temperature are flux functions,
\begin{eqnarray}\label{eq:Maxwellian}
F_{\spe 0} = \frac{n_\spe(\psi,t)}{(2\pi {T}(\psi,t))^{3/2}}
\exp\left(-\frac{u^2/2 + \mu B}{{T}(\psi,t)}\right).
\end{eqnarray}
 {Note the different notation employed for the density of the
Maxwellian, $n_\spe$, and for the total density, $N_\spe$, in
\eq{eq:defdensity}.}
The temperatures of all species are equal because due to our ordering,
$\tau_E$ is much larger than the collision time of all species,
including electrons. In particular, we are not expanding in
$\sqrt{m_e/m_i}  {\ll 1}$.  {To see the
  effect that such an expansion would have on the result, we consider
  the equations in \cite{Parra2015}, valid only for $B_p/B \ll 1$. By
  exploiting that $\sqrt{m_e/m_i} \ll 1$, reference \cite{Parra2015}
  finds momentum transport driven by the temperature difference
  between electrons and ions, $T_e - T_i$. Our equations will have
  this effect, although only in the limit $|T_e - T_i| \ll T_e \simeq
  T_i$. The temperature difference $T_e - T_i$ is contained in the
  piece $F_{\spe 2}$.}

The notation for the expansion of the
electrostatic potential has been given in \eq{eq:potentiallw1},
\eq{eq:potentialsw1},  {\eq{eq:notationphisw}, \eq{eq:potentiallw2} and \eq{eq:potentiallw3}.} As for the expansion
of the collision operator, we define each coefficient by
\begin{eqnarray}
  C_{\spe\spe'} = \epsilon_\spe C_{\spe\spe'}^{(1)} 
+ \epsilon_\spe^2 C_{\spe\spe'}^{(2)} + O(\epsilon_\spe^3).
\end{eqnarray}
Here, we have taken into account that the $O(\epsilon_\spe^0)$ term
equals zero because the collision operator vanishes when acting on
Maxwellians with the same temperatures.

Observe that $F_\spe(\bR,u,\mu,\theta)$ is obtained from the
gyrokinetic Fokker-Planck equation, and is therefore expressed
naturally in gyrokinetic coordinates, but the collision operator and
other functions entering the integrals in \eq{eq:Pinondim} are written
in coordinates $(\boldr,\bv)$. The simplest way to express
\eq{eq:Pinondim} is given by transforming the relevant pieces of
$F_\spe(\bR,u,\mu,\theta)$ to coordinates $(\boldr,\bv)$. In
\ref{sec:PiecesTransfFeuclideanCoor} we write the necessary
 {transformations. We} employ the notation of subsection
\ref{sec:gyrokinCoorTransf}. In
\ref{sec:ManipulationsPiecesCollOperators}, we manipulate the terms of
\eq{eq:Pinondim} containing collision operators. In
\ref{sec:FinalExpressionPiNonDim}, we give the final expression for
the radial flux of toroidal angular momentum.

%%%%%%%%%%%%%%%%%%%%%%%%%%%%%%
\subsubsection{Some pieces of the transformation of $F_\spe$ to
  euclidean coordinates. }
\label{sec:PiecesTransfFeuclideanCoor}
%%%%%%%%%%%%%%%%%%%%%%%%%%%%%%

We need the long-wavelength component of the transformation of the
Maxwellian to first and second order. To $O(\epsilon_\spe)$ the
calculation is given in \ref{sec:pullback} and the result is
\begin{eqnarray}\label{eq:pullback_order1_Maxwellian_text}
\fl \left[\cT^{-1*}_{\spe,1} F_{\spe 0}\right]^\lw &=& \cT^{-1*}_{\spe,0}F_{\spe 0}
+
\frac{\epsilon_\spe}{T}
\Bigg[
\bv\cdot{\mathbf V}^p_\spe +
\left(
\frac{v^2}{2T_{\spe}}-\frac{5}{2}
\right)\bv\cdot{\mathbf V}^T
\nonumber\\[5pt]
\fl&+&
\frac{Z_\spe}{B}
\bv\cdot(\bun\times\nabla_\boldr\varphi_0)
\Bigg]\cT^{-1*}_{\spe,0}F_{\spe 0},
\end{eqnarray}
where
\begin{equation}\label{eq:velocitiesPandT_text}
{\mathbf V}^p_\spe := \frac{1}{n_\spe B}\bun\times  {\nabla_\boldr} p_\spe, \quad
{\mathbf V}^T := \frac{1}{B}\bun\times  {\nabla_\boldr} {T},
\end{equation}
 {To $O(\epsilon_\spe^2)$,} the transformation is much more complicated. It
was computed in \cite{CalvoParra2012},
\begin{eqnarray} \label{T2F0}
\fl \left [ \mathcal{T}^{-1\ast}_{\spe, 2} F_{\spe 0} \right ]^\lw
= \frac{1}{2B^2} (\bv \times \bun) (\bv \times \bun) : \Bigg [
\nabla_\boldr \nabla_\boldr \ln n_\spe + 
 {\frac{Z_\spe}{{T}}} \nabla_\boldr
\nabla_\boldr \varphi_0
\nonumber\\\fl\hspace{0.5cm}
 - 
 {\frac{Z_\spe}{{T}^2}} ( \nabla_\boldr
\varphi_0 \nabla_\boldr {T} 
+ \nabla_\boldr {T} \nabla_\boldr
\varphi_0 ) + \left ( \frac{v^2}{2{T}} - \frac{3}{2} \right )
\nabla_\boldr \nabla_\boldr \ln {T}
\nonumber\\\fl\hspace{0.5cm} - \frac{v^2}{2{T}^3} 
\nabla_\boldr {T}
\nabla_\boldr
 {T} \Bigg ] \cT_{\spe,0}^{-1*}F_{\spe 0} 
+ \frac{1}{2B^2} \Bigg
[ (\bv \times \bun) \cdot \Bigg 
( \frac{\nabla_\boldr n_\spe}{n_\spe}
\nonumber\\\fl\hspace{0.5cm} +
 {\frac{Z_\spe\nabla_\boldr \varphi_0}{{T}}}
 + \left (
\frac{v^2}{2{T}} - \frac{3}{2} \right ) \frac{\nabla_\boldr
{T}}{{T}} \Bigg ) \Bigg ]^2 \cT_{\spe,0}^{-1*}F_{\spe 0} \nonumber\\\fl\hspace{0.5cm}+
\bR_{02}^\lw \cdot \left ( \frac{\nabla_\boldr n_\spe}{n_\spe} +
 {\frac{Z_\spe \nabla_\boldr \varphi_0}{{T}}} 
+ \left (
\frac{v^2}{2{T}} - \frac{3}{2} \right ) \frac{\nabla_\boldr
{T}}{{T}} \right ) \cT_{\spe,0}^{-1*}F_{\spe 0} 
\nonumber\\\fl\hspace{0.5cm}+
\frac{Z_\spe^4\tau_\spe^2}{2{T}^2}
\mathcal{T}^{-1\ast}_{\spe,0}\left[
(\phiwig_{\spe 1}^\sw)^2 \right]^\lw
 {\mathcal{T}^{-1\ast}_{\spe,0}} F_{\spe 0} 
+ \frac{1}{{T}} \Bigg [ \frac{Z_\spe^2\tau_\spe}{B} (\bv \times
\bun) \cdot \nabla_\boldr \varphi_1^\lw
\nonumber\\\fl\hspace{0.5cm}
 + Z_\spe^4\tau_\spe^2
\mathcal{T}^{-1\ast}_{\spe,0} \Psi_{\phi,\spe}^\lw 
+
Z_\spe^2\tau_\spe \mathcal{T}^{-1\ast}_{\spe,0} \Psi_{\phi B,\spe}^{\lw}
\nonumber\\\fl\hspace{0.5cm}
+ \mathcal{T}^{-1\ast}_{\spe,0} \Psi_{B,\spe} + 
 {\frac{Z_\spe
v_\bot^2}{4B^2} }(\matI - \bun \bun): \nabla_\boldr \nabla_\boldr \varphi_0
\nonumber\\\fl\hspace{0.5cm}+ \frac{Z_\spe^4\tau_\spe^2}{B}
\mathcal{T}^{-1\ast}_{\spe,0}\left[
{\phiwig_{\spe 1}^\sw}\partial_\mu 
\langle  {\phi_{\spe 1}^\sw}
\rangle
\right]^\lw
\Bigg ] \cT_{\spe,0}^{-1*}F_{\spe 0}.
\end{eqnarray}
Here,
\begin{eqnarray}\label{eq:R02lw}
  \fl \bR_{02}^\lw = \frac{1}{B} \Bigg[ \left ( v_{||} \bun +
    \frac{1}{4} \bv_\bot \right ) \bv \times \bun
  \nonumber\\
  \fl\hspace{1cm}
  + \bv \times \bun
  \left ( v_{||} \bun + \frac{1}{4} \bv_\bot \right ) \Bigg]
  \dotcross \nabla_\boldr \left ( \frac{\bun}{B} \right )
 \nonumber\\
\fl\hspace{1cm}+
  \frac{v_{||}}{B^2} \bv_\bot \cdot \nabla_\boldr \bun + \frac{v_{||}}{B^2}
  \bun \bun \cdot \nabla_\boldr \bun \cdot \bv_\bot \nonumber\\\fl\hspace{1cm}+
  \frac{\bun}{8B^2} [ \bv_\bot \bv_\bot - (\bv_\bot \times \bun)
  (\bv_\bot \times \bun) ]: \nabla_\boldr \bun \nonumber\\\fl\hspace{1cm}+
  \frac{v_\bot^2}{2B^3} \bun \bun \cdot \nabla_\boldr B -
  \frac{v_\bot^2}{4B^3} \nabla_{\boldr_\bot} B,
\end{eqnarray}
 {where we have used the convention $\mathbf{a}\mathbf{b}\dotcross\matM
  = \mathbf{a}\times(\mathbf{b}\cdot\matM)$,}
\begin{eqnarray} \label{eq:PsincphiB}
\fl \Psi^\lw_{\phi B,\spe} = - \frac{3 \mu}{2 Z_\spe\tau_\spe B^2}
\nabla_\bR B \cdot \nabla_{\bR} \varphi_0
- \frac{u^2}{Z_\spe\tau_\spe B^2} (\bun \cdot \nabla_\bR \bun)
 \cdot \nabla_{\bR}
\varphi_0,
\end{eqnarray}
\begin{eqnarray} \label{eq:Psincphi}
\fl \Psi^\lw_{\phi,\spe} = 
 - \frac{1}{2Z_\spe^2\tau_\spe^2 B^2} |\nabla_{\bR}
\varphi_0|^2 - \frac{1}{2B}\partial_\mu
\left[\langle (\phiwig_{\spe 1}^\sw)^2 \rangle
\right]^\lw,
\end{eqnarray}
and $\Psi_{B,\spe}$ is given in \eq{Psi2_B}.

We also need the long-wavelength component of the action of
$\cT_{\sigma,1}^{-1 *}$ on $F_{\spe 1}^\lw$ and $F_{\spe
  1}^\sw$. Employing repeatedly the results of
\ref{sec:pullback}, it is
straightforward to find that
\begin{eqnarray}\label{eq:T1F1lw}
\fl [\cT_{\sigma,1}^{-1 *} F_{\spe 1}^\lw]^\lw =  -
\cT_{\spe,0}^{-1 *}\Bigg\{
\rhobf\cdot\nabla_\bR\nonumber\\[5pt]
\fl\hspace{0.5cm}
 +
\Bigg(u\bun\cdot\nabla_\bR\bun\cdot\rhobf
+\frac{B}{4}[\rhobf(\rhobf\times\bun)+(\rhobf\times\bun)\rhobf]:\nabla_\bR\bun
\nonumber\\[5pt]
\fl\hspace{0.5cm}
-\mu\bun\cdot\nabla_\bR\times\bun
\Bigg)\partial_u
+
\Bigg(-\frac{\mu}{B}\rhobf\cdot\nabla_\bR B
\nonumber\\[5pt]
\fl\hspace{0.5cm}
-\frac{u}{4}[\rhobf(\rhobf\times\bun)+(\rhobf\times\bun)\rhobf]:\nabla_\bR\bun
+\frac{u\mu}{B}\bun\cdot\nabla_\bR\times\bun \nonumber\\[5pt]
\fl\hspace{0.5cm}
- \frac{u^2}{B}\bun\cdot\nabla_\bR\bun\cdot\rhobf
- \frac{Z_\spe}{B}\rhobf\cdot\nabla_\bR\varphi_0
\Bigg)\partial_\mu
\Bigg\}
F_{\spe 1}^\lw
\end{eqnarray}
and
\begin{eqnarray}\label{eq:T1F1sw}
\fl [\cT_{\sigma,1}^{-1 *} F_{\sigma 1}^\sw]^\lw =
- \cT_{\spe,0}^{-1 *}
\Bigg[
\Bigg( \frac{Z_\spe^2\tau_\spe}{B^2}
 (\nabla_{\bR_\perp/\epsilon_\spe}
\Phiwig_{\sigma 1}^\sw \times \bun ) \cdot \nabla_{\bR_\perp/\epsilon_\spe}
\nonumber\\[5pt]
\fl\hspace{0.5cm}
-
\frac{Z_\spe^2\tau_\spe \phiwig_{\spe 1}^\sw}{B}
\partial_\mu
\Bigg) F_{\spe 1}^\sw
\Bigg]^\lw.
\end{eqnarray}

The short-wavelength components of the action of $\cT_{\spe}^{-1*}$ on
$F_\spe$ to first and second order, denoted respectively by
$(\cT_{\spe}^{-1*} F_\spe)_1^\sw$ and $(\cT_{\spe}^{-1*}
F_\spe)_2^\sw$, are also needed. We introduce an operator $\modTinv$,
whose action on a phase-space function $G(\bR,u,\mu,\theta)$ is given
by
\begin{equation}\label{eq:auxiliaryOperator}
\fl\modTinv G(\boldr,\bv) :=
G\left(\boldr-\frac{\epsilon_\spe}{B(\boldr)}\bun(\boldr)\times\bv,
\bv\cdot\bun(\boldr), \frac{v_\bot^2}{2B(\boldr)},\arctan
\left(
\frac{\bv\cdot\eun_2(\boldr)}{\bv\cdot\eun_1(\boldr)}
\right)\right).
\end{equation}
This operator is useful to write some expressions involving the
short-wavelength pieces of the distribution functions and the
potential, for which it is not possible to Taylor expand the
dependence on $\boldr -\epsilon_\spe
B(\boldr)^{-1}\bun(\boldr)\times\bv$ around $\boldr$. Usually,
$G$ in \eq{eq:auxiliaryOperator} also depends slowly on $\bR$  {due to the quantities related to the magnetic field such as $B(\bR)$ or $\bun(\bR)$.} In $\modTinv G(\boldr,\bv)$ this dependence is
Taylor expanded around $\boldr$.

With the
help of \eq{eq:auxiliaryOperator} and \ref{sec:pullback} we obtain
\begin{eqnarray}\label{eq:TinvF1sw}
\fl
(\cT_{\spe}^{-1*} F_\spe)_1^\sw
=
\mathbb{T}_{\spe,0}F_{\spe 1}^\sw - \frac{Z_\spe^2\tau_\spe}{{T}}
\mathbb{T}_{\spe,0}\tilde\phi_{\spe 1}^\sw \cT_{\spe,0}^{-1*}F_{\spe 0}.
\end{eqnarray}

Finally, we have to compute $(\cT_{\spe}^{-1*} F_\spe)_2^\sw$. For
this, we need again the expressions in \ref{sec:pullback} and the
calculation of the short-wavelength transformation of the Maxwellian
to second order, explained in
\ref{sec:SecondOrderTransfMaxwellianSW}. The result is
\begin{eqnarray}\label{eq:TinvF2sw}
\fl
(\cT_\spe^{-1*}F_\spe)_2^\sw
=
\modTinv F_{\spe 2}^\sw
\nonumber\\[5pt]
\fl\hspace{0.5cm}
+
\Big[
\Big(\bR_{02}\cdot\modTinv\nabla_{\bR/\epsilon_\spe}
-(\cT_{\spe,0}^{-1*}\hat\mu_{\spe,1}^\lw 
+\modTinv \hat\mu_{\spe,1}^\sw)\modTinv\partial_\mu
\nonumber\\[5pt]
\fl\hspace{0.5cm}
-\cT_{\spe,0}^{-1*}\hat u_{\spe,1}^\lw\modTinv\partial_u
-(
\cT_{\spe,0}^{-1*}\hat\theta_{\spe,1}^\lw
+
\modTinv\hat\theta_{\spe,1}^\sw)\modTinv\partial_\theta
\Big)
F_{\spe 1}^\sw
\Big]^\sw
\nonumber\\[5pt]
\fl\hspace{0.5cm}
-
\mathbb{T}_{\spe,0}\hat\mu_{\spe,1}^\sw
\cT_{\spe,0}^{-1*}\partial_\mu F_{\spe 1}^\lw
+
\frac{[H_{01}^2]^\sw}{2 {T}^2}\cT_{\spe, 0}^{-1*} F_{\spe 0}
-\frac{H_{02}^\sw}{{T}}
\cT_{\spe, 0}^{-1*} F_{\spe 0}
\nonumber\\[5pt]
\fl\hspace{0.5cm}
-\frac{H_{01}^\sw}{{T} B}(\bv\times\bun)\cdot
\Bigg (
\frac{\nabla_\boldr n_\spe}{n_\spe} + \left ( \frac{v^2}{2{T}} -
\frac{5}{2} \right ) \frac{\nabla_\boldr {T}}{{T}}
\Bigg )
\cT_{\spe, 0}^{-1*} F_{\spe 0}
\nonumber\\[5pt]
\fl\hspace{0.5cm}
+\bR_{02}^\sw
\cdot
\Bigg (
\frac{\nabla_\boldr n_\spe}{n_\spe} + \left ( \frac{v^2}{2{T}} -
\frac{3}{2} \right ) \frac{\nabla_\boldr {T}}{{T}} \Bigg )
\cT_{\spe, 0}^{-1*} F_{\spe 0},
\end{eqnarray}
The corrections $\bR_{02}^\lw$,  {$\bR_{02}^\sw$, $H_{01}^\lw$, $H_{01}^\sw$} and $H_{02}^\sw$ are provided
in \eq{eq:R02lw}, \eq{eq:R02sw}, \eq{eq:H01lw}, \eq{eq:H01sw} and
\eq{eq:H02sw}. Using the expressions for $\hat u_{\spe,1}$,
$\hat \mu_{\spe, 1}$ and $\hat\theta_{\spe,1}$ in
\eq{eq:totalchangecoorfirstorderCorrections}, it is easy to obtain
\begin{eqnarray}\label{eq:hatu1pullback}
\fl \cT_{\spe, 0}^{-1*}\hat{u}_{\spe,1}^\lw &= 
-\frac{u}{B}\bun\cdot\nabla_\boldr\bun\cdot\bv\times\bun
-\frac{1}{4B}[
(\bv_\perp\times\bun)\bv_\perp + \bv_\perp(\bv_\perp\times\bun)
]
:\nabla_{\boldr}\bun
\nonumber\\[5pt]
\fl&-\mu\bun\cdot\nabla_\boldr\times\bun,
\end{eqnarray}
\begin{eqnarray}\label{eq:pullbackhatmulw}
\fl
\cT_{\spe,0}^{-1*}\hat{\mu}_{\spe,1}^\lw &=
\frac{\mu}{B^2}(\bv\times\bun)\cdot\nabla_{\boldr}B
+\frac{u}{4 B^2}
\left(
(\bv_\perp\times\bun)\bv_\perp + \bv_\perp(\bv_\perp\times\bun)
\right):\nabla_{\boldr}\bun
\nonumber\\[5pt]
\fl&+\frac{u\mu}{B}\bun\cdot\nabla_\boldr\times\bun
+\frac{u^2}{B^2}\bun\cdot\nabla_\boldr\bun\cdot(\bv\times\bun)
+\frac{Z_\spe}{B^2}(\bv\times\bun)
\cdot\nabla_\boldr\varphi_{0},
\end{eqnarray}
\begin{eqnarray}\label{eq:hattheta1pullbacklw}
\fl
\cT_{\spe,0}^{-1*}
\hat{\theta}_{\spe,1}^\lw&= 
\frac{1}{B}\bv_\perp
\cdot
\Bigg(
\nabla_\boldr\ln B + \frac{u^2}{2\mu B}\bun\cdot\nabla_\boldr\bun
\nonumber\\[5pt]
\fl&-\bun\times\nabla_\boldr\eun_2\cdot\eun_1
\Bigg)
-\frac{u}{8\mu B^2}
\left(
(\bv_\perp\times\bun)(\bv_\perp\times\bun)
 - \bv_\perp\bv_\perp
\right):\nabla_\boldr\bun
\nonumber\\[5pt]
\fl&
+\frac{u}{2B^2}\bun\cdot\nabla_\boldr B
+\frac{Z_\spe}{2\mu B^2}\bv_\perp\cdot\nabla_\boldr\varphi_0,
\end{eqnarray}
\begin{eqnarray}\label{eq:pullbackhatmusw}
\fl
\mathbb{T}_{\spe,0}\hat{\mu}_{\spe,1}^\sw(\boldr,t) &=
-\frac{Z_\spe^2\tau_\spe}{B(\boldr)}
\mathbb{T}_{\spe,0}\tilde\phi_{\spe 1}^\sw(\boldr,t)
\end{eqnarray}
and
\begin{eqnarray}\label{eq:hattheta1pullbacksw}
  \fl
  \mathbb{T}_{\spe,0}
  \hat{\theta}_{\spe,1}^\sw(\boldr,t) &= 
  \frac{Z_\spe^2\tau_\spe}{B(\boldr)}\mathbb{T}_{\spe,0}
\partial_\mu\tilde\Phi_{\spe 1}^\sw(\boldr,t).
\end{eqnarray}

%%%%%%%%%%%%%%%%%%%%%%%%%%%%%%
\subsubsection{Manipulations of terms in \eq{eq:Pinondim} that contain
  collision operators. }
\label{sec:ManipulationsPiecesCollOperators}
%%%%%%%%%%%%%%%%%%%%%%%%%%%%%%

We are ready to go back to \eq{eq:Pinondim} and write more explicitly
the terms on the right side. The term before last in \eq{eq:Pinondim} is
\begin{eqnarray}
\fl
-
\frac{\epsilon_s^2 }{6Z_\spe^2\tau_\spe^2}
\frac{1}{V'}
\partial_\psi
\left\langle
V'
\int 
\sum_{\spe'}C_{\spe\spe'}^\lw
R^3 (\bv\cdot\hat{\zetabf})^3
\dd^3 v
\right\rangle_\psi
=
\nonumber\\[5pt]
\fl\hspace{0.5cm}
-
\frac{\epsilon_s^3 }{6Z_\spe^3\tau_\spe^3}
\frac{1}{V'}
\partial_\psi
\left\langle
V'
\int 
\sum_{\spe'}C_{\spe\spe'}^{(1)\lw}
R^3 (\bv\cdot\hat{\zetabf})^3
\dd^3 v
\right\rangle_\psi,
\end{eqnarray}
where
\begin{eqnarray}\label{eq:C1lw}
\fl
C_{\spe\spe'}^{(1)\lw}=
C_{\spe\spe'}\Bigg[ \frac{1}{T} \left( \bv\cdot{\mathbf
V}^p_\spe + \left( \frac{v^2}{2T}-\frac{5}{2}
\right)\bv\cdot  {{\mathbf V}_\spe^T}
\right)\cT^{-1*}_{\spe,0}F_{\spe 0}
\nonumber\\[5pt]
\fl\hspace{1cm}
+ \cT^{-1*}_{\spe,0}F_{\spe
1}^\lw ,\cT^{-1*}_{\spe',0}F_{\spe' 0} \Bigg]
 +
\frac{Z_\spe\tau_\spe}{Z_{\spe'}\tau_{\spe'}}
C_{\spe\spe'} \Bigg[ \cT^{-1*}_{\spe,0}F_{\spe 0}, \frac{1}{
{T}} \Bigg( \bv\cdot{\mathbf V}^p_{\spe'}
\nonumber\\[5pt]
\fl\hspace{1cm}
 + \left(
\frac{v^2}{2{T}}-\frac{5}{2} \right)\bv\cdot  {{\mathbf
V}_{\spe'}^T}
\Bigg)\cT^{-1*}_{\spe',0}F_{\spe' 0} +\,
\cT^{-1*}_{\spe',0}F_{\spe' 1}^\lw
\Bigg].
\end{eqnarray}
Here, we have used \eq{eq:pullback_order1_Maxwellian_text} and the
Galilean invariance of the collision operator to drop the term
containing $\varphi_0$ in \eq{eq:pullback_order1_Maxwellian_text}.

The fourth term on the right side of \eq{eq:Pinondim} has two
contributions of different orders,
\begin{eqnarray}\label{eq:Pifourthterm}
\fl
-
\frac{\epsilon_s}{2Z_\spe \tau_\spe}
\left\langle
\int 
\sum_{\spe'}C_{\spe\spe'}^\lw
R^2 (\bv\cdot\hat{\zetabf})^2
\dd^3 v
\right\rangle_\psi
=
\nonumber\\[5pt]
\fl\hspace{0.5cm}
-
\frac{\epsilon_s^2}{2Z_\spe^2\tau_\spe^2}
\left\langle
\int 
\sum_{\spe'}C_{\spe\spe'}^{(1)\lw}
R^2 (\bv\cdot\hat{\zetabf})^2
\dd^3 v
\right\rangle_\psi
\nonumber\\[5pt]
\fl\hspace{0.5cm}
-
\frac{\epsilon_s^3}{2Z_\spe^3\tau_\spe^3}
\left\langle
\int 
\sum_{\spe'}C_{\spe\spe'}^{(2)\lw}
R^2 (\bv\cdot\hat{\zetabf})^2
\dd^3 v
\right\rangle_\psi.
\end{eqnarray}
The first-term on the right side can be computed by using
\eq{eq:C1lw}. As for the second term, employing
\eq{eq:pullback_order1_Maxwellian_text}, \eq{T2F0}, \eq{eq:T1F1lw},
\eq{eq:T1F1sw}, and \eq{eq:TinvF1sw} we immediately get
\begin{eqnarray}\label{eq:C2lw}
\fl  C_{\spe \spe'}^{(2)\lw} &=
C_{\sigma \sigma^\prime}
 \Big[\cT^{-1*}_{\spe,0} F_{\sigma 2}^\lw +
[\cT_{\sigma,1}^{-1 *} F_{\sigma 1}^\lw]^\lw 
 \nonumber\\
\fl&
+
[\cT_{\sigma,1}^{-1 *} F_{\sigma 1}^\sw]^\lw +
[\cT_{\sigma,2}^{-1 *} F_{\spe 0}]^\lw,
\cT^{-1*}_{\spe',0}F_{\spe' 0} \Big] \nonumber\\
\fl&+\left(\frac{Z_\spe\tau_\spe}{Z_{\spe'}\tau_{\spe'}}\right)^2 
C_{\sigma \sigma^\prime}
\Big[
\cT^{-1*}_{\spe,0}F_{\spe 0},\cT^{-1*}_{\spe',0} F_{\sigma' 2}^\lw
 \nonumber\\
\fl&
+
[\cT_{\sigma',1}^{-1 *} F_{\sigma' 1}^\lw]^\lw +
[\cT_{\sigma',1}^{-1 *} F_{\sigma' 1}^\sw]^\lw +
[\cT_{\sigma',2}^{-1 *} F_{\spe' 0}]^\lw
\Big] \nonumber\\
\fl& +
\frac{Z_\spe\tau_\spe}{Z_{\spe'}\tau_{\spe'}}
C_{\sigma\sigma^\prime}
\left[\cT^{-1*}_{\spe,0} F_{\sigma 1}^\lw
+ [\cT_{\sigma,1}^{-1 *} F_{\sigma 0}]^\lw,
\cT^{-1*}_{\spe',0}F_{\sigma^\prime 1}^\lw +
[\cT_{\sigma',1}^{-1 *} F_{\sigma' 0}]^\lw
\right ] \nonumber\\
\fl& +
\frac{Z_\spe\tau_\spe}{Z_{\spe'}\tau_{\spe'}}
\Bigg[
C_{\sigma\sigma^\prime} \Bigg[
\modTinv F_{\spe 1}^\sw -
\frac{Z_\spe^2\tau_\spe
}{{T}}
\modTinv \phiwig_{\spe 1}^\sw\cT_{\spe,0}^{-1 *}F_{\spe 0}
, 
\modTinvprime F_{\spe' 1}^\sw
\nonumber\\[5pt]
\fl&
-
\frac{Z_{\spe'}^2\tau_{\spe'} 
}{{T}}
\modTinvprime\phiwig_{\spe' 1}^\sw\cT_{\spe',0}^{-1 *}F_{\spe' 0}
\Bigg] \Bigg]^\lw.
\end{eqnarray}

%%%%%%%%%%%%%%%%%%%%%%%%%%%%%%
\subsubsection{Final expression for the radial flux of toroidal angular momentum. }
\label{sec:FinalExpressionPiNonDim}
%%%%%%%%%%%%%%%%%%%%%%%%%%%%%%

We turn to the first two terms on the right-hand side of
\eq{eq:Pinondim}. The second one is simple,
\begin{eqnarray}\label{eq:Pisecondterm}
\fl
\frac{\epsilon_s^2 }{2Z_\spe\tau_\spe^2}
\frac{1}{V'}\partial_\psi
\left\langle
V'
\partial_{\zeta/\epsilon_s}
\varphi
R^2 (\hat{\zetabf}\cdot\matP_\spe\cdot\hat{\zetabf})
\right\rangle_\psi
^\lw = 
\nonumber\\[5pt]
\fl\hspace{0.5cm}
\frac{\epsilon_s^3 }{2Z_\spe^2\tau_\spe^3}
\frac{1}{V'}\partial_\psi
\left[
\left\langle
V'
\partial_{\zeta/\epsilon_s}
\varphi_1^\sw
R^2 (\hat{\zetabf}\cdot\matP_{\spe 1}^\sw\cdot\hat{\zetabf})
\right\rangle_\psi
\right]^\lw,
\end{eqnarray}
because only the short-wavelength, $O(\epsilon_s)$ piece of the stress
tensor contributes. Namely,
\begin{eqnarray}
\fl
\matP_{\spe 1}^\sw (\boldr,t)
:=
\int \bv\bv (\cT_\spe^{-1*} F_\spe)^\sw_1(\boldr,\bv,t)\dd^3 v,
\end{eqnarray}
where $(\cT_\spe^{-1*} F_\spe)^\sw_1(\boldr,\bv,t)$ is defined in
\eq{eq:TinvF1sw}.

The first term on the right-hand side of \eq{eq:Pinondim} can be
rewritten as
\begin{eqnarray}\label{eq:Pifirstterm}
\fl
\frac{\epsilon_s^2}{\tau_\spe^2}
\left\langle
\partial_{\zeta/\epsilon_s}
\varphi
R  {N_\spe} (\bV_\spe\cdot\hat{\zetabf})
\right\rangle_\psi
^\lw = 
\nonumber\\[5pt]
\fl\hspace{0.5cm}
\frac{\epsilon_s^2}{\tau_\spe^2}
\left\langle
 {\partial_{\zeta/\epsilon_s}}
\varphi_1^\sw
R  {(N_\spe \bV_\spe)_1^\sw \cdot\hat{\zetabf}}
\right\rangle_\psi
^\lw
\nonumber\\[5pt]
\fl\hspace{0.5cm}
+
\frac{\epsilon_s^3}{Z_\spe\tau_\spe^3}
\left\langle
\partial_{\zeta/\epsilon_s}
\varphi_1^\sw
R  {(N_\spe \bV_\spe)_2^\sw \cdot\hat{\zetabf}}
\right\rangle_\psi
^\lw
\nonumber\\[5pt]
\fl\hspace{0.5cm}
+
\frac{\epsilon_s^3}{\tau_\spe^2}
\left\langle
\partial_{\zeta/\epsilon_s}
\varphi_2^\sw
R  {(N_\spe \bV_\spe)_1^\sw\cdot\hat{\zetabf}}
\right\rangle_\psi
^\lw
\end{eqnarray}
and therefore expressed in terms of
\begin{eqnarray}
 {(N_\spe \bV_\spe)_j^\sw} (\boldr,t)
:=
\tau_\spe
\int \bv (\cT_\spe^{-1*} F_\spe)^\sw_j(\boldr,\bv,t)\dd^3 v
\end{eqnarray}
for $j=1,2$. The quantity $(\cT_{\spe}^{-1*} F_\spe)_1^\sw$ has been
given in \eq{eq:TinvF1sw} and $(\cT_{\spe}^{-1*} F_\spe)_2^\sw$ has
been given in \eq{eq:TinvF2sw}.

We have found that the radial flux of toroidal angular momentum is
\begin{eqnarray}\label{eq:Pilw}
\Pi^\lw
=
\epsilon_s^2\Pi^\lw_{2} + \epsilon_s^3\Pi^\lw_{3} + O(\epsilon_s^4),
\end{eqnarray}
where
\begin{eqnarray}\label{eq:Pilw2}
\fl\Pi^\lw_{2}
=-
\sum_\spe
\frac{1}{2Z_\spe^2\tau_\spe^2}
\left\langle
\int 
\sum_{\spe'}C_{\spe\spe'}^{(1)\lw}
R^2 (\bv\cdot\hat{\zetabf})^2
\dd^3 v
\right\rangle_\psi
\nonumber\\[5pt]
\fl\hspace{0.5cm}
+\sum_\spe\frac{1}{\tau_\spe^2}
\left[
\left\langle
\partial_{\zeta/\epsilon_s}
\varphi_1^\sw
R  {(N_\spe \bV_\spe)_1^\sw\cdot\hat{\zetabf}}
\right\rangle_\psi
\right]^\lw
\end{eqnarray}
and
\begin{eqnarray}\label{eq:Pilw3}
\fl\Pi^\lw_{3}
=\sum_\spe\Bigg\{
-
\frac{1}{6Z_\spe^3\tau_\spe^3}
\frac{1}{V'}
\partial_\psi
\left\langle
V'
\int 
\sum_{\spe'}C_{\spe\spe'}^{(1)\lw}
R^3 (\bv\cdot\hat{\zetabf})^3
\dd^3 v
\right\rangle_\psi
\nonumber\\[5pt]
\fl\hspace{0.5cm}
-
\frac{1}{2Z_\spe^3\tau_\spe^3}
\left\langle
\int 
\sum_{\spe'}C_{\spe\spe'}^{(2)\lw}
R^2 (\bv\cdot\hat{\zetabf})^2
\dd^3 v
\right\rangle_\psi
+
\frac{1}{2Z_\spe\tau_\spe^2}\langle R^2 \rangle_\psi
\partial_{\epsilon_s^2 t} p_\spe
\nonumber\\[5pt]
\fl\hspace{0.5cm}
+\frac{1}{2Z_\spe^2\tau_\spe^3}
\frac{1}{V'}\partial_\psi
\left\langle
V'
\partial_{\zeta/\epsilon_s}
\varphi_1^\sw
R^2 (\hat{\zetabf}\cdot\matP_{\spe 1}^\sw\cdot\hat{\zetabf})
\right\rangle_\psi
^\lw
\nonumber\\[5pt]
\fl\hspace{0.5cm}
+
\frac{1}{Z_\spe\tau_\spe^3}
\left\langle
\partial_{\zeta/\epsilon_s}
\varphi_1^\sw
R  {(N_\spe \bV_\spe)_2^\sw\cdot\hat{\zetabf}}
\right\rangle_\psi
^\lw
\nonumber\\[5pt]
\fl\hspace{0.5cm}
+\frac{1}{\tau_\spe^2}
\left\langle
\partial_{\zeta/\epsilon_s}
\varphi_2^\sw
R  {(N_\spe \bV_\spe)_1^\sw\cdot\hat{\zetabf}}
\right\rangle_\psi
^\lw
\nonumber\\[5pt]
\fl\hspace{0.5cm}
-
\frac{1}{2Z_\spe\tau_\spe^2}
\left\langle
\int 
S_\spe
R^2 (\bv\cdot\hat{\zetabf})^2
\dd^3 v
\right\rangle_\psi
\Bigg\}.
\end{eqnarray}
The computation of the third term on the right-hand side of
\eq{eq:Pilw3} indicates that the transport equations for $n_\spe$ and
${T}$ are  {needed. To} calculate $n_\spe$, we use the density transport equation of
each species. Since we are assuming that the temperatures of all
species are equal, we can use the transport equation for the total
energy to determine ${T}$. The required transport equations were
calculated in \cite{CalvoParra2012} and can be found in subsection
\ref{sec:transportequations} of the present paper. Importantly,
 {$\langle F_{\spe 2}^\lw\rangle$} does not enter these transport
equations. The piece  {$\langle F_{\spe 2}^\lw\rangle$} only enters the
expression for $\Pi^\lw$ through $C_{\spe\spe'}^{(2)\lw}$, and it is
clear that adding a term $(Z_\spe^3 \tau_\spe^2 \varphi_2^\lw /{T})
F_{\spe 0}$ to  {$\langle F_{\spe 2}^\lw\rangle$} in those terms does not
change the result. This is why the $O(\epsilon_s^2)$ long-wavelength
quasineutrality equation is not needed.

Even though, in principle, $\Pi^\lw_2$ dominates \eq{eq:Pilw}, this
term is small in up-down symmetric tokamaks due to a symmetry of the
gyrokinetic equation, and it is comparable to
$\Pi_3^\lw$~\cite{parra11c, sugama11, Parra2015}.  {The
  feasibility of using external poloidal field coils to break up-down
  symmetry and generate intrinsic rotation has been explored in
  \cite{Camenen2009a, Camenen2009b, Ball2014}. The largest second
  order momentum flux obtained from up-down asymmetry has been
   {$\Pi_2^\lw/Q \sim 0.1$, where $Q$ is the turbulent heat flux normalized by the gyroBohm heat flux.} According to \cite{Parra2015, Parra2012},
   {$\Pi_3^\lw/Q \sim B/B_p$.} These estimates suggest that $\Pi_3^\lw$
  dominates over the up-down asymmetry contribution to $\Pi_2^\lw$ for
  $(B/B_p) \epsilon \gtrsim 0.1$. For $(B/B_p) \epsilon \lesssim 0.1$,
  the importance of $\Pi_3^\lw$ depends on how up-down asymmetric the
  tokamak is. Note that according to \cite{Ball2014} not all possible
  up-down asymmetric shapes are equally effective.}
    
    The next sections
  are devoted to present the equations that give the short and
  long-wavelength components of the distribution function and
  electrostatic potential that are needed to evaluate the right sides
  of \eq{eq:Pilw2} and \eq{eq:Pilw3}.

%%%%%%%%%%%%%%%%%%%%%%%%%%%%%%%%%%%%%%%%%%%%%%%%%%%%%%%%%%%%%%%
\section{Short-wavelength equations}
\label{sec:SWequations}
%%%%%%%%%%%%%%%%%%%%%%%%%%%%%%%%%%%%%%%%%%%%%%%%%%%%%%%%%%%%%%%

In this section the equations that determine the short-wavelength
components of the distribution functions and electrostatic potential
up to $O(\epsilon_s^2)$ are provided. The equations to $O(\epsilon_s)$
constitute the standard set of gyrokinetic equations. We give them for
completeness in subsection \ref{sec:FPandQuasineutSWfirstOrder}. The
equations for the $O(\epsilon_s^2)$ turbulent pieces are one of the
main results of this paper and are derived in subsections
\ref{sec:shortWaveFokkerPlanckEq} and
\ref{sec:shortWavePoissonEq}. Note that the $O(\epsilon_s^2)$
short-wavelength equations given in \cite{Parra2015} are valid
only if $B/B_p\gg 1$.

%%%%%%%%%%%%%%%%%%%%%%%%%%%%%%%%%%%%%%%%%%%%%%%%%%%%%%%%%%%%%%%
\subsection{Short-wavelength Fokker-Planck and quasineutrality
  equations to first order}
\label{sec:FPandQuasineutSWfirstOrder}
%%%%%%%%%%%%%%%%%%%%%%%%%%%%%%%%%%%%%%%%%%%%%%%%%%%%%%%%%%%%%%%

The first-order, short-wavelength terms of the Fokker-Planck equation
are
\begin{eqnarray}\label{eq:sworder1distfunction}
\fl\frac{1}{\tau_\spe}\partial_t F_{\spe 1}^\sw
+\left(u\bun\cdot\nabla_\bR-\mu\bun\cdot\nabla_\bR B\partial_u\right)
F_{\spe 1}^\sw
\nonumber\\[5pt]
\fl\hspace{0.5cm}
+
\left[
\frac{Z_\spe^2\tau_\spe}{B}
\left(\bun\times\nabla_{\bR_\perp/\epsilon_\spe}
\langle
\phi_{\spe 1}^\sw
\rangle
\right)
\cdot\nabla_{\bR_\perp/\epsilon_\spe}F_{\spe 1}^\sw
\right]^\sw
\nonumber\\[5pt]
\fl\hspace{0.5cm}
+\left(
\frac{u^2}{B}\bun\times\kappabf
+\frac{\mu}{B}\bun\times\nabla_\bR B
+\frac{Z_\spe}{B}\bun\times\nabla_\bR\varphi_0
\right)
\cdot\nabla_{\bR_\perp/\epsilon_\spe}F_{\spe 1}^\sw
\nonumber\\[5pt]
\fl\hspace{0.5cm}
+\frac{Z_\spe^2\tau_\spe}{B}
\left(\bun\times\nabla_{\bR_\bot/\epsilon_\spe}
\langle
\phi_{\spe 1}^\sw
\rangle
\right)
\cdot\nabla_\bR F_{\spe 0}
\nonumber\\[5pt]
\fl\hspace{0.5cm}
-Z_\spe^2\tau_\spe
\Big(
\bun\cdot\nabla_\bR\langle\phi_{\spe 1}^\sw\rangle
+\frac{u}{B}
\bun\times(\bun\cdot\nabla_\bR\bun)
\cdot
\nabla_{\bR_\perp/\epsilon_\spe}
\langle\phi_{\spe 1}^\sw\rangle
\Big)
\partial_u F_{\spe 0}
\nonumber\\[5pt]
\fl\hspace{0.5cm}
=
\sum_{\spe^\prime}
\left\langle \cT_{NP, \spe}^*  C_{\sigma \sigma^\prime}^{(1)\sw}
\right\rangle
.
\end{eqnarray}
Here we have used that $F_{\spe 1}$ does not depend on the gyrophase
and
\begin{eqnarray}\label{eq:C1sw}
\fl  C_{\sigma \sigma^\prime}^{(1)\sw} = C_{\sigma \sigma^\prime}
\left [\modTinv F_{\sigma 1}^\sw -
\frac{Z_\spe^2\tau_\spe}{{T}}
\modTinv\tilde\phi_{\spe 1}^\sw \cT_{\spe,0}^{-1
*} F_{\spe 0}, \cT_{\spe',0}^{-1
*}F_{\spe' 0}
 \right ]\nonumber\\[5pt]
\fl\hspace{0.5cm} + \
\frac{Z_\spe\tau_\spe}{Z_{\spe'}\tau_{\spe'}}C_{\sigma
\sigma^\prime} \left [ \cT_{\spe,0}^{-1 *}F_{\spe 0} ,
\modTinvprime F_{\sigma' 1}^\sw -
\frac{Z_{\spe'}^2\tau_{\spe'}}{{T}}
\modTinvprime\tilde\phi_{\spe'
1}^\sw \cT_{\spe',0}^{-1 *}F_{\spe' 0}
 \right ],
\end{eqnarray}
where we have employed \eq{eq:TinvF1sw}. The
transformation $(\boldr,\bv) = \cT_{NP,\spe}(\bR,u,\mu,\theta)$ is
defined in \eq{eq:changeNonPert}.

As for the short-wavelength,
first-order quasineutrality equation, we have
\begin{eqnarray}\label{eq:quasinautralitySWorder1appendix}
\fl\sum_\spe
&\int
B
\Bigg[-Z_\spe^2 \phiwig_{\spe
      1}^\sw \left(\boldr
-\epsilon_\spe\rhobf(\boldr,\mu,\theta),\mu,\theta,t\right)
\frac{F_{\spe 0}(\boldr,u,\mu,t)}{T_{\spe}(\boldr,t)}\nonumber\\[5pt]
\fl&
 +  \frac{1}{\tau_\spe} F_{\spe 1}^\sw
\left(\boldr
-\epsilon_\spe\rhobf(\boldr,\mu,\theta),u,\mu,t
\right)
\Bigg]
  \dd u \dd \mu \dd \theta = 0.
\end{eqnarray}

These equations are well known and were derived in this form in
reference \cite{CalvoParra2012}. They determine the lowest-order
turbulent contributions to the fields, $F_{\spe 1}^\sw$ and
$\varphi_1^\sw$.

%%%%%%%%%%%%%%%%%%%%%%%%%%%%%%%%%%%%%%%%%%%%%%%%%%%%%%%%%%%%%%%
\subsection{Short-wavelength Fokker-Planck equation to second order}
\label{sec:shortWaveFokkerPlanckEq}
%%%%%%%%%%%%%%%%%%%%%%%%%%%%%%%%%%%%%%%%%%%%%%%%%%%%%%%%%%%%%%%

In this section and in Section \ref{sec:shortWavePoissonEq}, we compute
the $O(\epsilon_\spe^2)$ terms of the short-wavelength Fokker-Planck
and quasineutrality equations, respectively.

%%%%%%%%%%%%%%%%%%%%%%%%%%%%%%%%%%%%%%%%%%%%%%%%%%%%%%%%%%%%%%%
\subsubsection{Gyrophase-dependent component of $F_{\spe 2}^\sw$.}
\label{sec:F2swGyrophaseDependent}
%%%%%%%%%%%%%%%%%%%%%%%%%%%%%%%%%%%%%%%%%%%%%%%%%%%%%%%%%%%%%%%

The equation that determines $F_{\spe 2}^\sw - \langle F_{\spe
  2}^\sw\rangle$ comes from the gyrophase-dependent part of the
$O(\epsilon_\spe)$ terms in the Fokker-Planck equation
\eq{eq:FokkerPlancknondimgyro},
\begin{eqnarray}\label{eq:F2swGyrophaseDependent}
\fl
\partial_\theta \left(
F_{\spe 2}^\sw - \langle
  F_{\spe 2}^\sw\rangle
\right)
\nonumber\\[5pt]
\fl\hspace{0.5cm}
=-\frac{1}{B}
\sum_{\spe^\prime}
\left(
\cT_{NP, \spe}^*  C_{\sigma \sigma^\prime}^{(1)\sw}
-
\left\langle \cT_{NP, \spe}^*  C_{\sigma \sigma^\prime}^{(1)\sw}
\right\rangle
\right)
.
\end{eqnarray}

%%%%%%%%%%%%%%%%%%%%%%%%%%%%%%%%%%%%%%%%%%%%%%%%%%%%%%%%%%%%%%%
\subsubsection{Gyroaveraged component of $F_{\spe 2}^\sw$.}
\label{sec:F2swGyroaveraged}
%%%%%%%%%%%%%%%%%%%%%%%%%%%%%%%%%%%%%%%%%%%%%%%%%%%%%%%%%%%%%%%

This subsection is devoted to the manipulations leading to the final
form of the second-order, short-wavelength, gyroaveraged Fokker-Planck
equation.  Recall equation \eq{eq:FokkerPlancknondimgyro} and recall
that $\dot\bR$ and $\dot u$ are given in \ref{sec:eqsofmotion}. We
have to identify the pieces of these equations of motion that
contribute to the gyroaveraged second-order Fokker-Planck equation at
short wavelengths.

Equation \eq{eq:dRdt} can be written as $\dot\bR_\spe =
\dot\bR_{\spe,0} + \epsilon_\spe \dot\bR_{\spe,1} +
\epsilon_\spe^2\dot\bR_{\spe,2} + O(\epsilon_\spe^3)$, where
$\dot\bR_{\spe,0} = u\bun$. The first-order terms are
\begin{eqnarray}
\fl
\dot\bR_{\spe,1}
=
\frac{1}{B}\bun\times(u^2\kappabf + \mu \nabla_\bR B)
\nonumber\\[5pt]
\fl\hspace{1cm}
+
\frac{Z_\spe}{B}\bun\times
\left(
\nabla_\bR\varphi_0 + Z_\spe\tau_\spe
\nabla_{\bR_\perp/\epsilon_\spe}\langle\phi_{\spe 1}^\sw\rangle
\right),
\end{eqnarray}
where $\kappabf:= \bun\cdot\nabla_\bR\bun$. We remind the reader that
we are using the conventions \eq{eq:potentiallw1},
\eq{eq:potentialsw1} and \eq{eq:notationphisw}, and also the formulae
\eq{eq:potentiallw2} and \eq{eq:potentiallw3}. The contribution of the
second-order terms of \eq{eq:dRdt} is
\begin{eqnarray}
\fl
\dot\bR_{\spe,2}
=
\nonumber\\[5pt]
\fl\hspace{1cm}
-\frac{u\mu}{B}(\nabla_\bR\times\bK)_\perp - \frac{u^3}{B^2}
(\bun\times\kappabf) (\nabla_\bR\times\bun)\cdot\bun
\nonumber\\[5pt]
\fl\hspace{1cm}
+
\left(
Z_\spe^2\tau_\spe\partial_u\Psi_{\phi B,\spe} + \partial_u\Psi_{B,\spe}
\right)\bun
\nonumber\\[5pt]
\fl\hspace{1cm}
+
\frac{1}{B}\bun\times
\left(
Z_\spe^4\tau_\spe^2\nabla_{\bR_\perp/\epsilon_\spe}\Psi_{\phi,\spe}^\sw
+
Z_\spe^2\tau_\spe
\nabla_{\bR_\perp/\epsilon_\spe} \Psi_{\phi B,\spe}^\sw
\right)
\nonumber\\[5pt]
\fl\hspace{1cm}
-
\frac{u}{B^2}(\nabla_\bR\times\bun)\cdot\bun
\bun\times\left(\mu\nabla_\bR B + Z_\spe\nabla_\bR\varphi_0+
Z_\spe^2\tau_\spe \nabla_{\bR_\perp/\epsilon_\spe}\langle\phi_{\spe 1}^\sw\rangle
\right)
\nonumber\\[5pt]
\fl\hspace{1cm}
+
\frac{1}{B}
\bun\times\left(Z_\spe^2\tau_\spe\nabla_\bR\varphi_1^\lw +
Z_\spe^3\tau_\spe^2 \nabla_{\bR_\perp/\epsilon_\spe}\langle\phi_{\spe 2}^\sw\rangle
\right),
\end{eqnarray}
where $\bK$ is defined in \eq{eq:defvectorK}.

The terms corresponding to the equation of motion of the parallel
velocity given in \eq{eq:dudt} can be written as $\dot u_\spe =
\dot u_{\spe,0} + \epsilon_\spe \dot u_{\spe,1} +
\epsilon_\spe^2\dot u_{\spe,2} + O(\epsilon_\spe^3)$, where
\begin{eqnarray}
\fl
\dot u_{\spe,0} = -\mu \bun\cdot\nabla_\bR B
\end{eqnarray}
and
\begin{eqnarray}
\fl
\dot u_{\spe,1}
=
-
Z_\spe^2\tau_\spe\bun\cdot\nabla_\bR
\left(
\langle\phi_{\spe 1}^\sw\rangle
+\varphi_1^\lw
\right)
\nonumber\\[5pt]
\fl\hspace{1cm}
-
\frac{u}{B}
(\bun\times\kappabf)\cdot
\left(\mu\nabla_\bR B +
Z_\spe\nabla_\bR\varphi_0
+
Z_\spe^2\tau_\spe \nabla_{\bR_\perp/\epsilon_\spe}\langle\phi_{\spe 1}^\sw\rangle
\right).
\end{eqnarray}
As for the second-order terms, $\dot u_{\spe,2}$, only the
short-wavelength component contributes,
\begin{eqnarray}
\fl
\dot u_{\spe,2}^\sw
=
\nonumber\\[5pt]
\fl\hspace{1cm}
-Z_\spe^2\tau_\spe
\bun\cdot\nabla_\bR
\left(
Z_\spe\tau_\spe\langle\phi_{\spe 2}^\sw\rangle
+Z_\spe^2\tau_\spe\Psi_{\phi,\spe}^\sw+\Psi_{\phi B,\spe}^\sw
\right)
\nonumber\\[5pt]
\fl\hspace{1cm}
-Z_\spe^2\tau_\spe\frac{u}{B}
(\bun\times\kappabf)\cdot
\nabla_{\bR_\perp/\epsilon_\spe}
\left(
Z_\spe\tau_\spe\langle\phi_{\spe 2}^\sw\rangle
+Z_\spe^2\tau_\spe\Psi_{\phi,\spe}^\sw+\Psi_{\phi B,\spe}^\sw
\right)
\nonumber\\[5pt]
\fl\hspace{1cm}
+
\frac{Z_\spe^2\tau_\spe}{B}
\Bigg(\frac{u^2}{B}
(\nabla_\bR\times\bun)\cdot\bun(\bun\times\kappabf)
+\mu(\nabla_\bR\times\bK)_\perp
\Bigg)\cdot
\nabla_{\bR_\perp/\epsilon_\spe}\langle\phi_{\spe 1}^\sw\rangle.
\end{eqnarray}

Then, the $O(\epsilon_\spe^2)$, gyroaveraged terms of
\eq{eq:FokkerPlancknondimgyro} at short wavelengths are
\begin{eqnarray}\label{eq:gyroaveragedF2sw}
\fl
\partial_t \langle F_{\spe 2}^\sw\rangle
+
\tau_\spe(u\bun\cdot\nabla_\bR - \mu\bun\cdot\nabla_\bR B \partial_u)
\langle F_{\spe 2}^\sw\rangle
\nonumber\\[5pt]
\fl\hspace{0.5cm}
+
\left[
\tau_\spe
\dot\bR_{\spe,1}
\cdot\nabla_{\bR_\perp/\epsilon_\spe} \langle F_{\spe 2}^\sw\rangle
\right]^\sw +
\tau_\spe
\dot u_{\spe,0} \partial_u \langle F_{\spe 2}^\sw\rangle
\nonumber\\[5pt]
\fl\hspace{0.5cm}
 { + 
\tau_\spe
\dot\bR_{\spe,1}^\sw \cdot\nabla_\bR F_{\spe 1}^\lw}
\nonumber\\[5pt]
\fl\hspace{0.5cm}
+
\tau_\spe
\dot\bR_{\spe,2}^\sw \cdot\nabla_\bR F_{\spe 0}
+
\left[
\tau_\spe
\dot\bR_{\spe,2}
\cdot\nabla_{\bR_\perp/\epsilon_\spe} F_{\spe 1}^\sw
\right]^\sw
\nonumber\\[5pt]
\fl\hspace{0.5cm}
+
\left[
\tau_\spe
\dot u_{\spe,1} \partial_u F_{\spe 1}^\sw
\right]^\sw
+
\tau_\spe
\dot u_{\spe,1}^\sw \partial_u F_{\spe 1}^\lw
+
\tau_\spe
\dot u_{\spe,2}^\sw \partial_u F_{\spe 0}
=
\nonumber\\[5pt]
\fl\hspace{0.5cm}
\left\langle
\left[\tau_\spe
\sum_{\spe'}\cT_{\spe}^*
C_{\spe\spe'}[\cT_{\spe}^{-1*} F_\spe, \cT_{\spe' }^{-1*} F_{\spe'}]
\right]_2^\sw
\right\rangle,
\end{eqnarray}
where we have used that the right-hand sides of \eq{eq:dRdt}, \eq{eq:dudt},
and \eq{eq:dthetadt} are gyrophase-independent, and so are $F_{\spe 1}$
and $F_{\spe 0}$. The gyrophase independence of the
equations of motion is the reason why \eq{eq:gyroaveragedF2sw} does
not contain contributions from the equation of motion for $\theta$,
\eq{eq:dthetadt}.

Next, we deal with the collision terms in
\eq{eq:gyroaveragedF2sw}. First, we write them as
\begin{eqnarray}\label{eq:colltermsOrder2sw}
\fl\left\langle
\left[
\tau_\spe
\sum_{\spe'}\cT_{\spe}^*
C_{\spe\spe'}[\cT_{\spe}^{-1*} F_\spe, \cT_{\spe' }^{-1*} F_{\spe'}]
\right]^\sw_2
\right\rangle =
\nonumber\\[5pt]
\fl
\hspace{1cm}
\left\langle
\tau_\spe
\sum_{\spe'}\cT_{\spe,0}^*
C_{\spe\spe'}[\cT_{\spe,0}^{-1*} F_{\spe 0}, (\cT_{\spe' }^{-1*} F_{\spe'})_2^\sw]
\right\rangle
\nonumber\\[5pt]
\fl
\hspace{1cm}
+
\left\langle
\tau_\spe
\sum_{\spe'}\cT_{\spe, 0 }^*
C_{\spe\spe'}[(\cT_{\spe}^{-1*} F_\spe)_2^\sw, \cT_{\spe' 0 }^{-1*}
F_{\spe' 0}]
\right\rangle
\nonumber\\[5pt]
\fl
\hspace{1cm}
+
 {\left\langle
\left[
\tau_\spe
\sum_{\spe'}\cT_{\spe, 0}^*
C_{\spe\spe'}[(\cT_{\spe}^{-1*} F_\spe)_1, (\cT_{\spe' }^{-1*} F_{\spe'})_1]
\right]^\sw
\right\rangle}
\nonumber\\[5pt]
\fl
\hspace{1cm}
+
\left\langle
\left[
\tau_\spe
\sum_{\spe'}\cT_{\spe,1}^*
C_{\spe\spe'}[\cT_{\spe,0}^{-1*} F_{\spe 0}, (\cT_{\spe' }^{-1*} F_{\spe'})_1]
\right]^\sw
\right\rangle
\nonumber\\[5pt]
\fl
\hspace{1cm}
+
\left\langle
\left[
\tau_\spe
\sum_{\spe'}\cT_{\spe,1}^*
C_{\spe\spe'}[(\cT_{\spe}^{-1*} F_\spe)_1, \cT_{\spe',0 }^{-1*}
F_{\spe' 0}]
\right]^\sw
\right\rangle
.
\end{eqnarray}

Let us start by writing the short-wavelength component of the action
of $\cT_{\spe,1}^*$ on any phase-space function $f(\boldr,\bv)$:
\begin{eqnarray}\label{eq:T1fsw}
\fl[\cT_{\spe,1}^* f]^\sw=
\hat\mu_{\spe,1}^\sw\partial_\mu(\cT_{\spe,0}^* f^\lw)
+
\hat\theta_{\spe,1}^\sw\partial_\theta(\cT_{\spe,0}^* f^\lw)
\nonumber\\[5pt]
\fl
\hspace{1cm}
+
[\bR_{\spe,2}^\sw\cdot\nabla_{\bR_\perp/\epsilon_\spe}(\cT_{NP,\spe}^*
f^\sw)]^\sw
+
[\mu_{\spe,1}^\sw\partial_\mu(\cT_{NP,\spe}^*
f^\sw)]^\sw
\nonumber\\[5pt]
\fl
\hspace{1cm}
+
[\theta_{\spe,1}^\sw\partial_\theta(\cT_{NP,\spe}^* f^\sw)]^\sw
+
(
\bR_{\spe,2}^\lw\cdot\nabla_{\bR_\perp/\epsilon_\spe}
\nonumber\\[5pt]
\fl
\hspace{1cm}
+
u_{\spe,1}^\lw\partial_u+\mu_{\spe,1}^\lw\partial_\mu+\theta_{\spe,1}^\lw
\partial_\theta
)
\cT_{NP,\spe}^* f^\sw,
\end{eqnarray}
with $\hat{u}_{\spe,1}$, $\hat{\mu}_{\spe,1}$, $\hat{\theta}_{\spe,1}$
given in \eq{eq:totalchangecoorfirstorderCorrections}, and
$\bR_{\spe,2}$, $u_{\spe,1}$, $\mu_{\spe,1}$, $\theta_{\spe,1}$ given
in \eq{R2}, \eq{u1}, \eq{mu1}, \eq{theta1}.  {Note the difference between $(\mathbf{R}_{\spe,2},
  u_{\spe,1}, \mu_{\spe,1}, \theta_{\spe,1})$ and $(\rhobf, \hat u_{\spe,1}, \hat
  \mu_{\spe,1}, \hat \theta_{\spe,1})$. The former give the give the
  lowest order terms of the perturbative transformation defined
  in \eq{eq:defPertTransf}. The latter are the $O(\epsilon_\sigma)$
  terms of the complete transformation $\cT_\spe$ from gyrokinetic
  coordinates to euclidean coordinates.}

Equation \eq{eq:T1fsw} is useful to write the two last terms on the
right side of \eq{eq:colltermsOrder2sw},
\begin{eqnarray}\label{eq:colltermsOrder2swPiece3}
\fl
\left\langle
\left[
\tau_\spe
\sum_{\spe'}\cT_{\spe,1}^*
C_{\spe\spe'}[\cT_{\spe,0}^{-1*} F_{\spe 0}, (\cT_{\spe' }^{-1*} F_{\spe'})_1]
\right]^\sw
\right\rangle
\nonumber\\[5pt]
\fl
\hspace{1cm}
+
\left\langle
\left[
\tau_\spe
\sum_{\spe'}\cT_{\spe,1}^*
C_{\spe\spe'}[(\cT_{\spe}^{-1*} F_\spe)_1, \cT_{\spe',0 }^{-1*}
F_{\spe' 0}]
\right]^\sw
\right\rangle=
\nonumber\\[5pt]
\fl
\hspace{1cm}
\left\langle
\tau_\spe
\sum_{\spe'}(\hat\mu_{\spe,1}^\sw\partial_\mu
+
\hat\theta_{\spe,1}^\sw\partial_\theta
)
\cT_{\spe,0}^* C_{\spe\spe'}^{(1)\lw}
\right\rangle
\nonumber\\[5pt]
\fl
\hspace{1cm}
+
\left\langle
\left[\tau_\spe
\sum_{\spe'}(\bR_{\spe,2}^\sw\cdot\nabla_{\bR_\perp/\epsilon_\spe}
+\mu_{\spe,1}^\sw\partial_\mu
+\theta_{\spe,1}^\sw\partial_\theta
)
\cT_{NP,\spe}^* C_{\spe\spe'}^{(1)\sw}
\right]^\sw
\right\rangle
\nonumber\\[5pt]
\fl
\hspace{1cm}
+
\left\langle
\tau_\spe
\sum_{\spe'}(\bR_{\spe,2}^\lw\cdot\nabla_{\bR_\perp/\epsilon_\spe}
+u_{\spe,1}^\lw\partial_u
+\mu_{\spe,1}^\lw\partial_\mu
+\theta_{\spe,1}^\lw\partial_\theta
)
\cT_{NP,\spe}^* C_{\spe\spe'}^{(1)\sw}
\right\rangle,
\end{eqnarray}
where $C_{\spe\spe'}^{(1)\lw}$ and $C_{\spe\spe'}^{(1)\sw}$ have been
defined, respectively, in equations \eq{eq:C1lw} and \eq{eq:C1sw}.

Define, for convenience,
\begin{eqnarray}\label{eq:TinvF1lw}
(\cT_{\spe}^{-1*} F_{\spe})_1^\lw = \cT_{\spe,0 }^{-1*} F_{\spe 1}^\lw + 
\left[\cT^{-1*}_{\spe,1} F_{\spe 0}\right]^\lw,
\end{eqnarray}
where the last term is given in
\eq{eq:pullback_order1_Maxwellian_text}.  Using \eq{eq:TinvF1sw} and
\eq{eq:TinvF1lw} we can write the third term on the right side of
\eq{eq:colltermsOrder2sw} as the sum of three pieces:
\begin{eqnarray}\label{eq:colltermsOrder2swPiece2}
\fl
 {\left\langle
\left[
\tau_\spe
\sum_{\spe'}\cT_{\spe, 0}^*
C_{\spe\spe'}[(\cT_{\spe}^{-1*} F_\spe)_1, (\cT_{\spe' }^{-1*} F_{\spe'})_1]
\right]^\sw
\right\rangle} = 
\nonumber\\[5pt]
\fl
\hspace{1cm}
\left\langle
\tau_\spe
\sum_{\spe'}\cT_{\spe, 0}^*
C_{\spe\spe'}[(\cT_{\spe}^{-1*} F_\spe)_1^\sw, (\cT_{\spe' }^{-1*} F_{\spe'})_1^\lw]
\right\rangle
\nonumber\\[5pt]
\fl
\hspace{1cm}
+
\left\langle
\tau_\spe
\sum_{\spe'}\cT_{\spe, 0}^*
C_{\spe\spe'}[(\cT_{\spe}^{-1*} F_\spe)_1^\lw, (\cT_{\spe' }^{-1*} F_{\spe'})_1^\sw]
\right\rangle
\nonumber\\[5pt]
\fl
\hspace{1cm}
+
 {\left\langle
\left[
\tau_\spe
\sum_{\spe'}\cT_{\spe, 0}^*
C_{\spe\spe'}[(\cT_{\spe}^{-1*} F_\spe)_1^\sw, (\cT_{\spe' }^{-1*} F_{\spe'})_1^\sw]
\right]^\sw
\right\rangle
.}
\end{eqnarray}

In order to reach explicit expressions for the first two terms on the
right side of \eq{eq:colltermsOrder2sw}, one needs
$(\cT_{\spe}^{-1*}F_\spe)_2^\sw$, which is given in \eq{eq:TinvF2sw}.

%%%%%%%%%%%%%%%%%%%%%%%%%%%%%%%%%%%%%%%%%%%%%%%%%%%%%%%%%%%%%%%
\subsection{Short-wavelength quasineutrality equation to second order}
\label{sec:shortWavePoissonEq}
%%%%%%%%%%%%%%%%%%%%%%%%%%%%%%%%%%%%%%%%%%%%%%%%%%%%%%%%%%%%%%%

The effort made in previous sections immediately gives the
$O(\epsilon_s^2)$ short-wavelength contributions to
\eq{eq:Quasineutralitynondim} in terms of already calculated
quantities. Namely,
\begin{eqnarray}\label{eq:quasineutralityOrder2sw}
\sum_\spe \frac{1}{Z_\spe\tau_\spe^2}\int \,
(\cT_\spe^{-1*}F_\spe)_2^\sw \dd ^3 v = 0,
\end{eqnarray}
where $(\cT_\spe^{-1*}F_\spe)_2^\sw$ is given in \eq{eq:TinvF2sw}.

%%%%%%%%%%%%%%%%%%%%%%%%%%%%%%%%%%%%%%%%%%%%%%%%%%%%%%%%%%%%%%%
\section{Long-wavelength equations}
\label{sec:LongWave}
%%%%%%%%%%%%%%%%%%%%%%%%%%%%%%%%%%%%%%%%%%%%%%%%%%%%%%%%%%%%%%%

The results of this section are taken from reference
\cite{CalvoParra2012}. They are included in this paper because, as we
have seen in subsection \ref{sec:radialfluxtoroidalmomentum}, the
$O(\epsilon_s)$ and $O(\epsilon_s^2)$ long-wavelength pieces of the
distribution functions and the $O(\epsilon_s)$ long-wavelength pieces
of the electrostatic potential are needed to compute the radial flux
of toroidal angular momentum.

%%%%%%%%%%%%%%%%%%%%%%%%%%%%%%%%%%%%%%%%%%%%%%%%%%%%%%%%%%%%%%%%%%%%%%%%%%%%
\subsection{Long-wavelength Fokker-Planck equation up to second order}
\label{sec:LongWaveFokkerPlanckEq}
%%%%%%%%%%%%%%%%%%%%%%%%%%%%%%%%%%%%%%%%%%%%%%%%%%%%%%%%%%%%%%%

The equation for the first-order piece $F_{\spe}^\lw$ is conveniently
written in terms of
\begin{eqnarray}\label{eq:defGspe1}
\fl G_{\spe 1}^\lw&:=& F_{\spe 1}^\lw
+
\Bigg\{
\frac{Z_\spe^2\tau_\spe}{{T}}\varphi_1^\lw
+\frac{Iu}{B}
\left(
\frac{Z_\spe}{{T}}\partial_\psi\varphi_0
+\Upsilon_\spe
\right)
\Bigg\}F_{\spe 0},
\end{eqnarray}
where
\begin{eqnarray}\label{eq:defUpsilon}
\Upsilon_\spe := 
\partial_\psi \ln n_\spe+
\left(\frac{u^2/2 +\mu B}{{T}}-\frac{3}{2}\right)
\partial_\psi \ln {T}\, .
\end{eqnarray}
It reads
\begin{eqnarray}\label{eq:Vlasovorder1gyroav4}
\fl
\left(u\bun\cdot\nabla_\bR - \mu\bun\cdot\nabla_\bR
B\partial_u
\right)
G_{\spe 1}^\lw
\nonumber\\[5pt]
\fl\hspace{1cm}
= \sum_{\sigma'}\cT_{\spe,0}^*
C_{\spe\spe'}\left[ \cT^{-1*}_{\spe,0}\left(G_{\spe
1}^\lw-\frac{Iu}{B}
\Upsilon_\spe
 F_{\spe 0}\right)
,\cT^{-1*}_{\spe',0}F_{\spe' 0} \right]
\nonumber\\[5pt]
\fl\hspace{1cm} + \sum_{\sigma'}
\frac{Z_\spe\tau_\spe}{Z_{\spe'}\tau_{\spe'}}\cT_{\spe,0}^*
C_{\spe\spe'} \Bigg[  {\cT^{-1*}_{\spe,0}}F_{\spe 0},
 {\cT^{-1*}_{\spe',0}}\Bigg(G_{\spe' 1}^\lw- \frac{Iu}{B}
\Upsilon_{\spe'}
F_{{\spe'} 0}
\Bigg)
\Bigg].
\end{eqnarray}

In order to write the second-order, long-wavelength, gyroaveraged
Fokker-Planck equation, we define
\begin{eqnarray}\label{eq:defG2again3}
\fl G_{\spe 2}^\lw & =  {\langle F_{\sigma 2}^\lw \rangle}
+
\frac{Z_\spe^3\tau_\spe^2}{{T}}\varphi_2^\lw
F_{\spe 0}
+
\frac{Iu}{B}
\partial_\psi G_{\sigma 1}^\lw   -
\frac{Z_\spe^2\tau_\spe \varphi_1^\lw}{u} \partial_u
G_{\sigma 1}^\lw 
\nonumber\\
\fl & -
\frac{I}{B} \left ( \mu \partial_\psi B 
 + Z_\spe \partial_\psi\varphi_0  \right )
\partial_u G_{\sigma 1}^\lw 
- \frac{1}{2}
\left(
\frac{Z_\spe^2\tau_\spe\varphi_1^\lw}{{T}}
\right)^2
F_{\sigma 0}
 \nonumber\\\fl &
 - \frac{Z_\spe^2\tau_\spe I u}{{T} B}
\varphi_1^\lw F_{\spe 0} \Bigg(
\frac{Z_\spe}{{T}}
\partial_\psi\varphi_0 
+\Upsilon_\spe-
 \frac{1}{{T}} \partial_\psi
{T}  \Bigg)\nonumber\\
\fl &
-\frac{1}{2B^2}\left(
\left ({Iu} \right )^2+\mu B|\nabla_\bR\psi|^2
\right)
\Bigg[
- \frac{2Z_\spe}{{T}^2}
 \partial_\psi\varphi_0 
\partial_\psi {T}
\nonumber\\
\fl &
+ \Bigg(
\frac{Z_\spe}{{T}}
\partial_\psi\varphi_0  +\Upsilon_\spe
 \Bigg )^2
+
\partial_\psi^2\ln n_\spe
\nonumber\\
\fl & 
+
\left(\frac{u^2/2 +\mu B}{{T}}-\frac{3}{2}\right)
\partial_\psi^2\ln {T}
\nonumber\\
\fl & 
-
\frac{u^2/2 +\mu B}{{T}}
(\partial_\psi \ln {T})^2
 +
\frac{Z_\spe}{{T}}
\partial_\psi^2 \varphi_0
\Bigg]F_{\spe 0}
 \nonumber\\
\fl& + \left[ \frac{Z_\spe^2\tau_\spe}{
u \bB \cdot
\nabla_\bR \Theta} F_{\sigma 1}^\sw (\bun\times
\nabla_{\bR_\perp/\epsilon_\spe}
\langle \phi_{\spe 1}^\sw \rangle) \cdot \nabla_\bR
\Theta \right]^\lw\nonumber\\
\fl& -
\left \langle \frac{1}{ u \bun \cdot
  \nabla_\bR \Theta} \rhobf \cdot \nabla_\bR \Theta
\sum_{\sigma^\prime}\cT_{\spe,0}^*C_{\sigma
    \sigma^\prime}^{(1)\lw} \right \rangle
\nonumber\\
\fl
&- \frac{Z_\spe^4\tau_\spe^2}{2{T}^2}
\left[
\left \langle 
(\phiwig_{\spe 1}^\sw)^2 \right \rangle \right]^\lw F_{\spe
0}
+\mu\bun\cdot\nabla_\bR\times\bun
\left(
\frac{u}{B}\partial_\mu
-\partial_u 
\right)G_{\spe 1}^\lw
\nonumber\\
\fl&
+ 
\left\langle \cT_{\spe,0}^{*}
\left[\cT_{\spe,1}^{-1*}F_{\spe 1}^\lw\right]^\lw
\right\rangle
+
\left\langle \cT_{\spe,0}^{*}
\left[\cT_{\spe,2}^{-1*}F_{\spe 0}\right]^\lw
\right\rangle,
\end{eqnarray}
where the two last terms are computed in \eq{eq:T1F1lwAveraged} and
\eq{gyroaverageT2F0}.

Then, the equation is
\begin{eqnarray} \label{eq:eqH2sigma}
\fl\Big( u \bun \cdot &  \nabla_\bR  - \mu \bun \cdot \nabla_\bR
B\partial_u \Big) G_{\spe 2}^\lw +
Z_\spe^2\tau_\spe \partial_{\epsilon_s^2 t} F_{\spe 0}
\nonumber\\\fl & - \bun \cdot \nabla_\bR \Theta
\partial_\psi \Bigg \{
\frac{Z_\spe^2 \tau_\spe}{\bB \cdot \nabla_\bR
\Theta} \left[ F_{\sigma 1}^\sw
(\nabla_{\bR_\perp/\epsilon_\spe} \langle \phi_{\spe 1}^\sw
\rangle \times \bun) \cdot \nabla_\bR \psi
\right]^\lw \nonumber\\ \fl & +
 \frac{1}{\bun \cdot \nabla_\bR
\Theta} \left \langle
 \left(\frac{Iu}{B} +
\rhobf \cdot \nabla_\bR \psi \right ) \sum_{\sigma^\prime}
\cT_{\spe,0}^*C_{\sigma \sigma^\prime}^{(1)\lw} \right \rangle
\Bigg \} \nonumber\\
\fl &
- \partial_u \Bigg \{ \Bigg[
 Z_\spe^2\tau_\spe
 F_{\sigma 1}^\sw
 \Big(  \bun\cdot\nabla_\bR
\langle\phi_{\spe
 1}^\sw\rangle
\nonumber\\
\fl &
 + \frac{\mu}{u B} \partial_\Theta B(\bun \times
\nabla_\bR \Theta) \cdot
\nabla_{\bR_\perp/\epsilon_\spe}\langle\phi_{\spe 1}^\sw\rangle
 \nonumber\\
\fl &+ 
\frac{u}{B} [\bun \times (\bun \cdot
\nabla_\bR \bun)] \cdot
\nabla_{\bR_\perp/\epsilon_\spe}\langle\phi_{\spe 1}^\sw\rangle 
 \Big)  \Bigg]^\lw \nonumber\\
\fl & - \left \langle \frac{I}{B}
\left ( \mu \partial_\psi B  +
Z_\spe
 \partial_\psi\varphi_0  \right )
\sum_{\sigma^\prime} \cT_{\spe,0}^*C_{\sigma \sigma^\prime}^{(1)\lw} \right
\rangle \Bigg \} \nonumber\\\fl & + \partial_\mu \Bigg \langle
 \frac{1}{B}
\rhobf \cdot \nabla_\bR \psi \left ( \mu 
\partial_\psi B + Z_\spe \partial_\psi\varphi_0
 \right ) \sum_{\sigma^\prime}
\cT_{\spe,0}^*C_{\sigma \sigma^\prime}^{(1)\lw}
 \Bigg \rangle
\nonumber\\
\fl & = - \sum_{\sigma^\prime} \partial_u \Bigg \langle
 \Bigg [ \frac{Z_\spe^2 \tau_\spe
 \varphi_1^\lw}{u}
 +
\mu \bun \cdot \nabla_\bR \times \bun \nonumber\\
\fl & - \frac{1}{u}\rhobf
\cdot \left ( \mu \partial_\Theta B \nabla_{\bR} \Theta 
 + u^2 \bun \cdot \nabla_\bR \bun \right )
\Bigg ]  \cT_{\spe,0}^*C_{\sigma \sigma^\prime}^{(1)\lw}
 \Bigg \rangle  \nonumber\\
\fl &
+
\sum_{\sigma^\prime} \partial_\mu \Bigg \langle
\Bigg[ \frac{u \mu}{B} \bun \cdot \nabla_\bR
\times
\bun
\nonumber\\
\fl &
 - \frac{1}{B}\rhobf \cdot \left ( \mu \partial_\Theta B
 \nabla_{\bR} \Theta
 + u^2 \bun \cdot \nabla_\bR
\bun \right ) \Bigg ] \cT_{\spe,0}^*C_{\sigma \sigma^\prime}^{(1)\lw}
\Bigg\rangle
\nonumber\\
\fl &+\sum_{\sigma^\prime}\left[ \left\langle
\cT_{\spe, 1}^* C_{\sigma
\sigma^\prime}^{(1)\sw}\right\rangle
\right]^\lw
+\sum_{\sigma^\prime} \left \langle 
\cT_{\spe,0}^*C_{\sigma \sigma^\prime}^{(2)\lw}  \right
\rangle
\nonumber\\
\fl &
+ Z_\spe^2\tau_\spe^2\left\langle\cT_{\spe,0}^* S_\spe\right\rangle.
\end{eqnarray}
The term $\left \langle \cT_{\spe,0}^*C_{\sigma
    \sigma^\prime}^{(2)\lw} \right \rangle$ is given in \ref{sec:C2nc},
whereas the gyroaverage of $[\cT_{\spe, 1}^{*} C_{\spe
  \spe^\prime}^{(1)\sw}]^\lw$ was computed in \cite{CalvoParra2012}
and the result is
\begin{eqnarray}\label{eq:turbpiececollgyroave}
\fl
\Big\langle
\Big[ &\cT_{\spe,1}^*C_{\spe\spe'}^\sw
\Big]^\lw
\Big\rangle
=-
\partial_\mu
\Bigg\langle\Bigg[
\frac{Z_\spe^2\tau_\spe}{B}
\tilde\phi_{\spe 1}^\sw
\nonumber\\[5pt]
\fl&
\Bigg\{
\cT_{NP, \spe}^* C_{\sigma \sigma^\prime}
\left [\modTinv F_{\sigma 1}^\sw -
\frac{Z_\spe^2\tau_\spe}{{T}}
\modTinv\tilde\phi_{\spe 1}^\sw \cT_{\spe,0}^{-1
*} F_{\spe 0}, \cT_{\spe',0}^{-1
*}F_{\spe' 0}
 \right ]
\nonumber\\[5pt]
\fl& + \
\frac{Z_\spe\tau_\spe}{Z_{\spe'}\tau_{\spe'}}
\cT_{NP, \spe}^*
C_{\sigma
\sigma^\prime} \Bigg[ \cT_{\spe,0}^{-1 *}F_{\spe 0} ,
\modTinvprime F_{\sigma' 1}^\sw
\nonumber\\[5pt]
\fl&
 -
\frac{Z_{\spe'}^2\tau_{\spe'}}{{T}}
\modTinvprime\tilde\phi_{\spe'
1}^\sw \cT_{\spe',0}^{-1 *}F_{\spe' 0}
 \Bigg]
\Bigg\}
\Bigg]^\lw
\Bigg\rangle.
\end{eqnarray}
In equation \eq{eq:eqH2sigma} we have added the source term
 {$Z_\spe^2\tau_\spe^2\left\langle\cT_{\spe,0}^* S_\spe\right\rangle$
that} was not considered in reference \cite{CalvoParra2012}.

As explained in detail in subsection 5.1 of reference
\cite{CalvoParra2012}, the long-wavelength Fokker-Planck equation, at
any order in $\epsilon_\spe$, has a non-zero kernel. In order to fix
the component along the kernel, one has to impose some conditions. For instance, we can impose, for $j=1,2$,
\begin{eqnarray}\label{eq:gaugefixingExample}
\left\langle
\int B G_{\spe j}^\lw\dd u\dd\mu\dd\theta
\right\rangle_\psi
 = 0, \mbox{ for every $\spe$, and}
\nonumber\\
\left\langle\sum_\spe\frac{1}{Z_\spe^j\tau_\spe^j} \int B \left(u^2/2+\mu B\right)
G_{\spe j}^\lw\dd u\dd\mu\dd\theta
\right\rangle_\psi
 = 0.
\end{eqnarray}
This is a natural way to fix the ambiguity but there are infinitely
many different possibilities.

%%%%%%%%%%%%%%%%%%%%%%%%%%%%%%%%%%%%%%%%%%%%%%%%%%%%%%%%%%%%%%%
\subsection{Long-wavelength quasineutrality equation to first order}
\label{sec:longwaveQuasineutrality}
%%%%%%%%%%%%%%%%%%%%%%%%%%%%%%%%%%%%%%%%%%%%%%%%%%%%%%%%%%%%%%%

To order $\epsilon_s^0$, the quasineutrality equation simply gives a
relation among the densities of $F_{\spe 0}$,
\begin{eqnarray}\label{eq:gyroPoissonlw2order0}
\sum_\spe
Z_\spe n_\spe(\boldr,t) = 0.
\end{eqnarray}

In terms of the function $G_{\spe 1}^\lw$
defined in \eq{eq:defGspe1}, the
quasineutrality equation to first order gives
\begin{eqnarray}\label{eq:gyroPoissonlw2order1aux2}
\fl
\sum_\spe  \Bigg( \frac{1}{\tau_\spe}\int B(\boldr) G_{\spe
  1}^\lw(\boldr,u,\mu,t)\dd u\dd\mu\dd\theta
\nonumber\\[5pt]
\fl \hspace{1cm}-\frac{Z_\spe^2
  }{{T}}n_\spe(\boldr,t) \varphi_1^\lw(\boldr,t)\Bigg) =
0.
\end{eqnarray}

It is important to note that $\varphi_1^\lw$ can be determined from
the above equation only up to an additive function of $\psi$. This is
connected to the ambiguity in the determination of $F_{\spe 1}^\lw$
mentioned at the end of subsection
\ref{sec:LongWaveFokkerPlanckEq}. In other words, only
$\bun\cdot\nabla_\bR \varphi_1^\lw$ can be found. Without loss of
generality, we take
\begin{equation}
\left\langle
\varphi_1^\lw
\right\rangle_\psi = 0.
\end{equation}
Again, a longer discussion on this is given in subsection 5.1 of
reference \cite{CalvoParra2012}.

%%%%%%%%%%%%%%%%%%%%%%%%%%%%%%%%%%%%%%%%%%%%%%%%%%%%%%%%%%%%%%%%%%%%%%%%%%%%
\subsection{Transport equations for density and energy}
\label{sec:transportequations}
%%%%%%%%%%%%%%%%%%%%%%%%%%%%%%%%%%%%%%%%%%%%%%%%%%%%%%%%%%%%%%%

Momentum transport calculations require knowledge of the time
evolution of the  {functions of $\psi$} entering the Maxwellian, i.e. the density of
each species and the temperature, which is the same for all of
them. We take the relevant transport equations from
\cite{CalvoParra2012}. The equation for $n_\spe$ is
\begin{eqnarray}\label{eq:transportEq_density}
\fl
\partial_{\epsilon_s^2 t} n_{\spe}(\psi,t)
= \frac{1}{V'(\psi)}\partial_\psi
\Bigg\langle
V'(\psi)
\int\dd u\dd\mu\dd\theta\
 \Bigg \{
\nonumber\\[5pt]
\fl\hspace{1cm}
\left[ F_{\sigma 1}^\sw (\nabla_{\bR_\perp/\epsilon_\spe}
\langle \phi_{\spe 1}^\sw
\rangle \times \bun) \cdot \nabla_\bR \psi \right]^\lw
\nonumber\\[5pt]
\fl
\hspace{1cm}
+\frac{B}{Z_\spe^2 \tau_\spe}
  \left \langle
 \left(\frac{Iu}{B} +
\rhobf \cdot \nabla_\bR \psi \right ) \sum_{\sigma^\prime}
C_{\sigma \sigma^\prime}^{(1)\lw} \right \rangle
\Bigg \}\Bigg\rangle_\psi
\nonumber\\[5pt]
\fl
\hspace{1cm}
 {+ \left\langle \int B \left\langle\cT_{\spe,0}^* S_\spe\right\rangle \dd u\dd\mu\dd\theta \right \rangle_\psi
.}
\end{eqnarray}
As we have explained in subsection
\ref{sec:radialfluxtoroidalmomentum}, with these equations and the
transport equation for the total energy,
\begin{eqnarray}\label{eq:transportEq_totalEnergyAux}
  \fl \partial_{\epsilon_s^2 t} \left( \sum_\spe 
\frac{3}{2}n_\spe (\psi,t) {T} (\psi,t) \right) 
  =
  \nonumber\\
  \fl \hspace{0.5cm}
  \frac{1}{V^\prime(\psi)} \partial_\psi \Bigg\langle
  V^\prime (\psi) \int \left(u^2/2 + \mu B\right)
  \sum_\spe \Bigg\{
  \nonumber\\
  \fl \hspace{0.5cm}
 \left[ F_{\sigma 1}^\sw
    (\nabla_{\bR_\perp/\epsilon_\spe} \langle \phi_{\spe 1}^\sw
    \rangle \times \bun) \cdot \nabla_\bR \psi \right]^\lw
  \nonumber\\ \fl \hspace{0.5cm}
  + \frac{B}{Z_\spe^2 \tau_\spe} \left \langle
    \left(\frac{Iu}{B} +
      \rhobf \cdot \nabla_\bR \psi \right ) \sum_{\sigma^\prime}
    \cT_{\spe,0}^*C_{\sigma \sigma^\prime}^{(1)\lw} \right \rangle \Bigg\}
  \dd u\dd\mu\dd\theta\Bigg\rangle_\psi
  \nonumber\\
  \fl \hspace{0.5cm} -\Bigg\langle \sum_\spe
  \int B \Bigg [ F_{\sigma 1}^\sw
  \Big(  u \bun\cdot\nabla_\bR
  \langle\phi_{\spe
    1}^\sw\rangle
  \nonumber\\
  \fl \hspace{0.5cm}
  + \frac{\mu}{B} (\bun \times
  \nabla_\bR B) \cdot
  \nabla_{\bR_\perp/\epsilon_\spe}\langle\phi_{\spe 1}^\sw\rangle
  \nonumber\\
  \fl \hspace{0.5cm} + 
  \frac{u^2}{B} [\bun \times (\bun \cdot
  \nabla_\bR \bun)] \cdot
  \nabla_{\bR_\perp/\epsilon_\spe}\langle\phi_{\spe 1}^\sw\rangle 
  \Big)  \Bigg]^\lw \dd u \dd \mu \dd \theta \Bigg \rangle_\psi
  \nonumber\\
  \fl \hspace{0.5cm} +
  \Bigg\langle
  \sum_{\spe,\spe'} \frac{1}{Z_\spe^2\tau_\spe}
  \int
  B
  \left(u^2/2+\mu B\right)
   \left[\left\langle
      \cT_{\spe, 1}^* C_{\sigma
        \sigma^\prime}^{(1)\sw}
    \right\rangle
  \right]^\lw
  \dd u\dd\mu\dd\theta\Bigg\rangle_\psi
\nonumber\\[5pt]
\fl
\hspace{1cm}
 {+ \sum_\spe
\left\langle\int B\left(\frac{u^2}{2}+\mu B\right)
 \left\langle\cT_{\spe,0}^* S_\spe\right\rangle \dd u\dd\mu\dd\theta \right\rangle_\psi
,}
\end{eqnarray}
the transport equation for the temperature can be calculated.

 { Note that the calculation of density and energy transport
  only requires  {gyrokinetic equations correct to
  $O(\epsilon_s)$.} In other words, it is not necessary to know the
  long-wavelength radial electric field to solve the transport
  equations for $n_\spe$ and $T$. Higher order gyrokinetic equations
  are needed to determine the transport of toroidal angular
  momentum. This is in contrast to what happens in the high-flow
  ordering~\cite{Hinton1985, Artun1994, Brizard1995, Sugama1998,
    Abel2013}, where the plasma velocity is assumed to be $O(c_s)$. In
  that setting, density, energy and toroidal angular momentum
  transport can be calculated from the solution of the gyrokinetic
  equations to $O(\epsilon_s)$.  }

%%%%%%%%%%%%%%%%%%%%%%%%%%%%%%%%%%%%%%%%%%%%%%%%%%%%%%%%%%%%%%%
\section{Conclusions}
\label{sec:discussion}
%%%%%%%%%%%%%%%%%%%%%%%%%%%%%%%%%%%%%%%%%%%%%%%%%%%%%%%%%%%%%%%

Recently, it has been proven that the computation of radial transport
of toroidal angular momentum in a tokamak requires, in the low flow
regime, high-order gyrokinetics~\cite{ParraBarnesCalvoCatto2012}. This
issue has received much attention because the problem is equivalent to
determining the tokamak intrinsic rotation profile. In references
\cite{parra10a, parra11NF, Parra2015}, the equations needed to
calculate intrinsic rotation have been given under the assumption
$B/B_p \gg 1$, where $B_p$ is the magnitude of the poloidal magnetic
field.  However, a set of equations valid for tokamaks with large
$B_p$ was missing, and it has been derived in this paper. In this
section, we restrict ourselves to point out the equations in the text
that have to be solved by a code intended to calculate radial
transport of toroidal angular momentum in a tokamak. Denote by
$\epsilon_s\sim\rho_i/L$ the gyrokinetic expansion parameter, where
$\rho_i$ is the ion Larmor radius and $L$ is the variation length of
the magnetic field. One needs
\begin{itemize}
\item[]The Fokker-Planck equation at short-wavelengths up to
  $O(\epsilon_s^2)$, equations \eq{eq:sworder1distfunction},
  \eq{eq:F2swGyrophaseDependent} and \eq{eq:gyroaveragedF2sw};
\item[]The quasineutrality equation at short-wavelengths up to
  $O(\epsilon_s^2)$, equations \eq{eq:quasinautralitySWorder1appendix} and
  \eq{eq:quasineutralityOrder2sw};
\item[]The Fokker-Planck equation at long-wavelengths up to
  $O(\epsilon_s^2)$, equations \eq{eq:Vlasovorder1gyroav4} and
  \eq{eq:eqH2sigma};
\item[]The quasineutrality equation at long-wavelengths to
  $O(\epsilon_s)$, equation \eq{eq:gyroPoissonlw2order1aux2};
\item[]The transport equation for the density of each species $\spe$,
  \eq{eq:transportEq_density}, and for the total energy,
  \eq{eq:transportEq_totalEnergyAux};
\item[]The formula for the radial flux of toroidal angular momentum, equations \eq{eq:Pilw}, \eq{eq:Pilw2} and \eq{eq:Pilw3}.
\end{itemize}

%%%%%%%%%%%%%%%%%%%%%%%%%%%%%%%%%%%%%%%%%%

\ack I.~C. acknowledges the hospitality of Merton College and of the
Rudolf Peierls Centre for Theoretical Physics, University of Oxford,
where part of this work was completed. This work has been carried out
within the framework of the EUROfusion Consortium and has received
funding from the European Union's Horizon 2020 research and innovation
programme under grant agreement number 633053. The views and opinions
expressed herein do not necessarily reflect those of the European
Commission. This research was supported in part by grant
ENE2012-30832, Ministerio de Econom\'{\i}a y Competitividad, Spain, by
the RCUK Energy Programme (grant number EP/I501045) and by US DoE
grant DE-SC008435.

%%%%%%%%%%%%%%%%%%%%%%%%%%%%%%%%%%%%%%%%%%%%%%%%%%%%%%%%%%%%%%%%%%%%%%%%%%%
%%%%%%%%%%%%%%%%%%%%%%%%%%%%%%%%%%%%%%%%%%%%%%%%%%%%%%%%%%%%%%%%%%%%%%%%%%%
\appendix
%%%%%%%%%%%%%%%%%%%%%%%%%%%%%%%%%%%%%%%%%%%%%%%%%%%%%%%%%%%%%%%%%%%%%%%%%%%
%%%%%%%%%%%%%%%%%%%%%%%%%%%%%%%%%%%%%%%%%%%%%%%%%%%%%%%%%%%%%%%%%%%%%%%%%%%

%%%%%%%%%%%%%%%%%%%%%%%%%%%%%%%%%%%%%%%%
\section{Gyrokinetic equations of motion}
\label{sec:eqsofmotion}
%%%%%%%%%%%%%%%%%%%%%%%%%%%%%%%%%%%%%%%%

The gyrokinetic equations of motion derived in
\cite{ParraCalvo2011} are reproduced here,
\begin{eqnarray} \label{eq:dRdt}
\fl 
\dot\bR& = 
\nonumber \\
\fl &\left ( u + Z_\spe^2\tau_\spe
\epsilon^2_\spe \partial_u\Psi_{\phi B,\spe} +
\epsilon^2_\spe \partial_u\Psi_{B,\phi} \right )
\frac{\bB_\spe^*}{B_{||,\spe}^*}
\nonumber \\
\fl &
+ \frac{1}{B_{||,\spe}^*} \bun \times \Bigg (
\epsilon_\spe \mu \nabla_\bR B 
+ Z_\spe^2\tau_\spe \epsilon_\spe
\nabla_{\bR_\perp/ \epsilon_\spe}
 \langle\phi_\spe\rangle 
 \nonumber \\
\fl &
 +
{Z_\spe^4\tau_\spe^2 \epsilon^2_\spe}
 \nabla_{\bR_\bot/
\epsilon_\spe} \Psi_{\phi,\spe}
 + Z_\spe^2\tau_\spe \epsilon^2_\spe
\nabla_{\bR_\bot/ \epsilon_\spe}
 \Psi_{\phi B,\spe}\nonumber \\ \fl &
 +
Z_\spe^4\tau_\spe^2 \epsilon_\spe^3 \nabla_\bR \Psi_{\phi,\spe}
+ Z_\spe^2\tau_\spe \epsilon_\spe^3 \nabla_\bR \Psi_{\phi B,\spe}
 + \epsilon_\spe^3 \nabla_\bR \Psi_{B,\spe} \Bigg ),
\end{eqnarray}
\begin{eqnarray} \label{eq:dudt}
\fl
\dot u&=
\nonumber \\
\fl &
 - \frac{\mu}{B^*_{||,\spe}} \bB_\spe^*
\cdot \nabla_\bR B - Z_\spe^2\tau_\spe \epsilon_\spe \bun \cdot \nabla_\bR
\langle\phi_\spe\rangle  \nonumber \\
\fl &
- Z_\spe^4\tau_\spe^2 \epsilon_\spe^2
 \bun \cdot \nabla_\bR
\Psi_{\phi,\spe}
- Z_\spe^2\tau_\spe \epsilon_\spe^2
 \bun \cdot \nabla_\bR
\Psi_{\phi B,_\spe}
 \nonumber \\[3pt]
\fl &
- \epsilon_\spe^2 \bun \cdot
\nabla_\bR \Psi_{B,\spe}
 - \frac{1}{B_{||,\spe}^*} \Big[ u \bun \times (\bun
\cdot \nabla_\bR \bun)
 \nonumber \\[3pt]
\fl &
 - \epsilon_\spe \mu (\nabla_\bR \times \bK)_\bot
\Big] \cdot \Bigg ( Z_\spe^2\tau_\spe \epsilon_\spe
\nabla_{\bR_\bot/ \epsilon_\spe} \langle\phi_\spe\rangle
 \nonumber \\[3pt]
\fl & +
Z_\spe^4\tau_\spe^2 \epsilon_\spe^2
 \nabla_{\bR_\bot/
\epsilon_\spe} \Psi_{\phi,\spe}
 + Z_\spe^2\tau_\spe \epsilon_\spe^2
\nabla_{\bR_\bot/\epsilon_\spe} \Psi_{\phi B,_\spe}  \nonumber \\
\fl &
+
Z_\spe^4\tau_\spe^2 \epsilon_\spe^3 \nabla_\bR \Psi_{\phi,\spe}
+ Z_\spe^2\tau_\spe
\epsilon_\spe^3 \nabla_\bR \Psi_{\phi B,\spe}
 +
\epsilon_\spe^3 \nabla_\bR \Psi_{B,\spe} \Bigg) ,
\end{eqnarray}
\begin{eqnarray} \label{eq:dmudt}
\fl \dot\mu = 0,
\end{eqnarray}
\begin{eqnarray} \label{eq:dthetadt}
\fl
\dot\theta =
 -
\frac{1}{\epsilon_\spe} B + O(1)
.
\end{eqnarray}
Here,
\begin{equation} \label{Bstar}
\bB_\spe^* (\bR, u, \mu) := \bB(\bR) + \epsilon_\spe u \nabla_\bR \times
\bun (\bR)
 - \epsilon^2_\spe \mu \nabla_\bR \times \bK (\bR)
\end{equation}
and
\begin{equation}\label{eq:defvectorK}
\bK (\bR) := \frac{1}{2} \bun(\bR) \bun(\bR) \cdot \nabla_\bR
\times \bun(\bR) - \nabla_\bR \eun_2 (\bR) \cdot \eun_1 (\bR).
\end{equation}
 {The last term of \eq{eq:defvectorK} depends on the
  definition of the gyrophase, i.e. on the choice of $\eun_1 (\bR)$
  and $\eun_2 (\bR)$. It is common to say that this term is
  ``gyrogauge dependent''. However, only $\nabla_{\bR}\times \bK
  (\bR)$ enters the gyrokinetic equations, which was shown to be
  gyrogauge independent in \cite{Littlejohn1981}.}

The parallel component of $\bB_\spe^*$,
\begin{eqnarray} \label{Bstar_par}
B^*_{||,\spe} (\bR, u, \mu) := \bB_\spe^* (\bR, u, \mu) \cdot \bun (\bR)=
\nonumber \\[5pt]
\hspace{1cm}  B(\bR) + \epsilon_\spe u \bun(\bR) \cdot \nabla_\bR
\times \bun (\bR)
\nonumber \\[5pt]
\hspace{1cm}
 - \epsilon^2_\spe \mu \bun(\bR) \cdot \nabla_\bR
\times \bK (\bR),
\end{eqnarray}
is the Jacobian of the gyrokinetic transformation $\cT_\spe$ to
$O(\epsilon_\spe^2)$. Finally,
\begin{eqnarray}
\fl \Psi_{\phi,\spe} &= \frac{1}{2 B^2} \left
\langle \nabla_{(\bR_\bot/\epsilon_\spe)} \Phiwig_\spe \cdot \left
( \bun \times \nabla_{(\bR_\bot/\epsilon_\spe)} \phiwig_\spe
\right ) \right\rangle
\nonumber\\ \fl &
 - \frac{1}{2 B} \partial_\mu \langle
\phiwig^2_\spe \rangle, \label{Psi2_phi}
\end{eqnarray}
\begin{eqnarray}
\fl \Psi_{\phi B,\spe}& =  - \frac{u}{B} \left
\langle \left ( \nabla_{(\bR_\bot/\epsilon_\spe)} \phiwig_\spe
\times \bun \right ) \cdot \nabla_\bR \bun \cdot \rhobf \right
\rangle
\nonumber\\ \fl &
 - \frac{\mu}{2 B^2} \nabla_\bR B \cdot
\nabla_{(\bR_\bot/\epsilon_\spe)} \langle\phi_\spe\rangle
- \frac{1}{B} \nabla_\bR B
\cdot \langle \phiwig_\spe\, \rhobf \rangle
\nonumber\\ \fl & - \frac{1}{4 B} \left \langle
\nabla_{(\bR_\bot/\epsilon_\spe)} \phiwig_\spe \cdot \left [
\rhobf \rhobf - ( \rhobf \times \bun ) (\rhobf \times \bun) \right
] \cdot \nabla_\bR B \right \rangle  \nonumber\\ \fl &
- \frac{u^2}{B} \bun \cdot \nabla_\bR \bun \cdot
\left \langle \partial_\mu \phiwig_\spe\, \rhobf \right \rangle -
\frac{u^2}{2 \mu B} \bun \cdot \nabla_\bR \bun \cdot \langle
\phiwig_\spe\, \rhobf \rangle \nonumber\\\fl &
+
\frac{u}{4} \nabla_\bR \bun : \left \langle \partial_\mu
\phiwig_\spe\, \left [ \rhobf ( \rhobf \times \bun ) + (\rhobf
\times \bun) \rhobf \right ] \right \rangle \nonumber\\\fl & +
\frac{u}{4 \mu} \nabla_\bR \bun : \left \langle \phiwig_\spe\,
\left [ \rhobf ( \rhobf \times \bun ) + (\rhobf \times \bun)
\rhobf \right ] \right \rangle \label{Psi2_phiB}
\end{eqnarray}
and
\begin{eqnarray}
\fl \Psi_{B,\spe} &=  - \frac{3u^2 \mu}{2B^2} \bun \cdot \nabla_\bR
\bun \cdot \nabla_\bR B
\nonumber \\
\fl & 
 + \frac{\mu^2}{4B} (\matI - \bun \bun) :
\nabla_\bR \nabla_\bR \bB \cdot \bun
\nonumber \\
\fl &
 - \frac{3\mu^2}{4B^2}
|\nabla_{\bR_\bot} B|^2 
+ \frac{u^2 \mu}{2B}
\nabla_\bR \bun : \nabla_\bR \bun
\nonumber \\
\fl &  + \left(\frac{\mu^2}{8} -
\frac{u^2 \mu}{4B}\right) \nabla_{\bR_\perp} \bun : (\nabla_{\bR_\perp}
\bun)^\mathrm{T}
\nonumber \\
\fl & 
 - \left(\frac{3 u^2 \mu}{8B} +
\frac{\mu^2}{16}\right) (\nabla_\bR \cdot \bun)^2 \nonumber \\ \fl
& + \left(\frac{3 u^2 \mu}{2B}-\frac{u^4}{2B^2}\right) |\bun \cdot
\nabla_\bR \bun|^2 
\nonumber \\
\fl & 
+ \left(\frac{ u^2 \mu}{8B} -
\frac{\mu^2}{16}\right) (\bun \cdot \nabla_\bR \times \bun)^2.
\label{Psi2_B}
\end{eqnarray}
Here $\matrixtop{\mathbf{M}}^\mathrm{T}$ is the transpose of an
arbitrary matrix matrix $\matrixtop{\mathbf{M}}$ and
\begin{equation}
\Phiwig_\spe(\bR,\mu,\theta,t) :=
\int^\theta\phiwig_\spe(\bR,\mu,\theta',t)\dd\theta'\ ,
\end{equation}
where the lower limit of the integral is chosen such that
$\langle\Phiwig_\spe\rangle = 0$.

%%%%%%%%%%%%%%%%%%%%%%%%%%%%%%%%%%%%%%%%%%%%%%%%%%%%%%%%%%%%%%%
\section{Lowest-order terms of the perturbative transformation,
  $\cT_{P,\spe}$}
\label{sec:lowestordertermsTP}
%%%%%%%%%%%%%%%%%%%%%%%%%%%%%%%%%%%%%%%%%%%%%%%%%%%%%%%%%%%%%%%

The expressions for the corrections $\bR_{\spe,2}$, $u_{\spe,1}$,
$\mu_{\spe,1}$, and $\theta_{\spe,1}$ found in \cite{ParraCalvo2011},
and valid for arbitrary magnetic geometry, are
\begin{eqnarray}
\fl\bR_{\spe,2} &=& - \frac{2u}{B} \bun \bun \cdot \nabla_\bR \bun \cdot
(\rhobf \times \bun)- \frac{u}{B} \bun \times \nabla_\bR \bun \cdot \rhobf
\nonumber\\
\fl&&
 - \frac{1}{8} \bun \left [ \rhobf \rhobf -
(\rhobf \times \bun) (\rhobf \times \bun) \right ]: \nabla_\bR
\bun 
\nonumber\\
\fl&&- \frac{1}{2B} \rhobf \rhobf \cdot \nabla_\bR B
- \frac{Z_\spe^2 \tau_\spe}{B^2}
 \bun \times \nabla_{(\bR_\bot/\epsilon_\spe)} \Phiwig_\spe,
\label{R2}
\\[5pt]
\fl u_{\spe,1} &=& u \bun \cdot \nabla_\bR \bun \cdot \rhobf
\nonumber\\
\fl&&
 - \frac{B}{4}
\left [ \rhobf (\rhobf \times \bun) + (\rhobf \times \bun) \rhobf
\right ] : \nabla_\bR \bun, \label{u1}
\\[5pt]
\fl\mu_{\spe,1} &=& 
- \frac{Z_\spe^2\tau_\spe\phiwig_\spe}{B}
- \frac{u^2}{B} \bun \cdot \nabla_\bR \bun
\cdot \rhobf
\nonumber\\
\fl&&
 + \frac{u}{4} \left [ \rhobf (\rhobf \times \bun) +
(\rhobf \times \bun) \rhobf \right ]: \nabla_\bR \bun,
\label{mu1}
\\[5pt]
\fl\theta_{\spe,1} &=& \frac{Z_\spe^2 \tau_\spe}{B}
\partial_\mu \Phiwig_\spe
+\frac{u^2}{2\mu B} \bun \cdot \nabla_\bR \bun \cdot
(\rhobf \times \bun)
\nonumber\\
\fl&&
 + \frac{u}{8\mu} \left [ \rhobf \rhobf -
(\rhobf \times \bun) (\rhobf \times \bun) \right ]: \nabla_\bR
\bun
\nonumber\\
\fl&&+ \frac{1}{B} (\rhobf \times \bun) \cdot \nabla_\bR B
 .
\label{theta1}
\end{eqnarray}

%%%%%%%%%%%%%%%%%%%%%%%%%%%%%%%%%%%%%%%%%%%%%%%%%%%%%%%%%%%%%%%
\section{Gyrokinetic transformation, $\cT_\spe$,  to first order}
\label{sec:pullback}
%%%%%%%%%%%%%%%%%%%%%%%%%%%%%%%%%%%%%%%%%%%%%%%%%%%%%%%%%%%%%%%

In this appendix we provide explicit expressions for the gyrokinetic
transformation $(\boldr,\bv)=\cT_\spe(\bR,u,\mu,\theta,t)$ to
order $\epsilon_\spe$. Define
\begin{eqnarray}\label{eq:defcoordinates0}
v_{||}&:=\bv\cdot\bun(\boldr),\\[5pt]
\mu_0&:=\frac{(\bv-v_{||}\bun(\boldr))^2}{2B(\boldr)},\\[5pt]
\theta_0&:=\arctan
\left(
\frac{\bv\cdot\eun_2(\boldr)}{\bv\cdot\eun_1(\boldr)}
\right).
\end{eqnarray}
The result is
\begin{eqnarray}\label{eq:totalchangecoorfirstorder}
\fl \boldr &= \bR + \epsilon_\spe\rhobf
+ O(\epsilon_\spe^2)
,
\nonumber\\[5pt]
\fl v_{||} &= u
+ \epsilon_\spe\hat{u}_{\spe,1}
+ O(\epsilon_\spe^2),\nonumber\\[5pt]
\fl \mu_0 &= \mu + \epsilon_\spe\hat{\mu}_{\spe,1}
+ O(\epsilon_\spe^2),\nonumber\\[5pt]
\fl \theta_0 &= \theta
+\epsilon_\spe\hat{\theta}_{\spe,1}
+ O(\epsilon_\spe^2),
\end{eqnarray}
where
\begin{eqnarray}\label{eq:totalchangecoorfirstorderCorrections}
\fl \hat{u}_{\spe,1} &= 
u\bun\cdot\nabla_\bR\bun\cdot\rhobf+
\frac{B}{4}[\rhobf(\rhobf\times\bun)
+(\rhobf\times\bun)\rhobf]
:\nabla_{\bR}\bun
\nonumber\\[5pt]
\fl&-\mu\bun\cdot\nabla_\bR\times\bun,
\nonumber\\[5pt]
\fl\hat{\mu}_{\spe,1} &=
-\frac{\mu}{B}\rhobf\cdot\nabla_{\bR}B
-\frac{u}{4}
\left(
\rhobf(\rhobf\times\bun)
+(\rhobf\times\bun)\rhobf
\right):\nabla_{\bR}\bun
\nonumber\\[5pt]
\fl&+\frac{u\mu}{B}\bun\cdot\nabla_\bR\times\bun
-\frac{u^2}{B}\bun\cdot\nabla_\bR\bun\cdot\rhobf
-\frac{Z_\spe^2\tau_\spe}{B}\tilde\phi_{\spe 1},
\nonumber\\[5pt]
\fl\hat{\theta}_{\spe,1}&= 
(\rhobf\times\bun)
\cdot
\Bigg(
\nabla_\bR\ln B + \frac{u^2}{2\mu B}\bun\cdot\nabla_\bR\bun
\nonumber\\[5pt]
\fl&-\bun\times\nabla_\bR\eun_2\cdot\eun_1
\Bigg)
-\frac{u}{8\mu}
\left(
\rhobf\rhobf - (\rhobf\times\bun)(\rhobf\times\bun)
\right):\nabla_\bR\bun
\nonumber\\[5pt]
\fl&
+\frac{u}{2B^2}\bun\cdot\nabla_\bR B
+\frac{Z_\spe^2\tau_\spe}{B}\partial_\mu\tilde\Phi_{\spe 1}
.
\end{eqnarray}
 {Note} the difference between $(\mathbf{R}_{\spe,2},
  u_{\spe,1}, \mu_{\spe,1}, \theta_{\spe,1})$, defined in
  \ref{sec:lowestordertermsTP}, and $(\rhobf, \hat u_{\spe,1}, \hat
  \mu_{\spe,1}, \hat \theta_{\spe,1})$. The former give the give the
  lowest order terms of the perturbative transformation defined
  in \eq{eq:defPertTransf}. The latter are the $O(\epsilon_\sigma)$
  terms of the complete transformation $\cT_\spe$ from gyrokinetic
  coordinates to euclidean coordinates.

It is useful to have the long-wavelength limit of the previous
expressions at hand. Employing
\eq{eq:potentiallw2} and \eq{eq:potentiallw3}, we get
\begin{eqnarray}\label{eq:totalchangecoorfirstorderCorrectionsLW}
\fl  {\hat{u}_{\spe,1}^\lw} &= \hat{u}_{\spe,1}
\nonumber\\[5pt]
\fl\hat{\mu}_{\spe,1}^\lw &=
-\frac{\mu}{B}\rhobf\cdot\nabla_{\bR}B
-\frac{u}{4}
\left(
\rhobf(\rhobf\times\bun)
+(\rhobf\times\bun)\rhobf
\right):\nabla_{\bR}\bun
\nonumber\\[5pt]
\fl&+\frac{u\mu}{B}\bun\cdot\nabla_\bR\times\bun
-\frac{u^2}{B}\bun\cdot\nabla_\bR\bun\cdot\rhobf
-\frac{Z_\spe}{B}\rhobf\cdot\nabla_\bR\varphi_0,
\nonumber\\[5pt]
\fl\hat{\theta}_{\spe,1}^\lw&= 
(\rhobf\times\bun)
\cdot
\Bigg(
\nabla_\bR\ln B + \frac{u^2}{2\mu B}\bun\cdot\nabla_\bR\bun
\nonumber\\[5pt]
\fl&-\bun\times\nabla_\bR\eun_2\cdot\eun_1
\Bigg)
-\frac{u}{8\mu}
\left(
\rhobf\rhobf - (\rhobf\times\bun)(\rhobf\times\bun)
\right):\nabla_\bR\bun
\nonumber\\[5pt]
\fl&
+\frac{u}{2B^2}\bun\cdot\nabla_\bR B
+\frac{Z_\spe}{2\mu B}\left(\rhobf\times\bun\right)\cdot\nabla_\bR\varphi_0
.
\end{eqnarray}

 {To write \eq{eq:pullback_order1_Maxwellian_text} in Section
\ref{sec:radialfluxtoroidalmomentum}, we need to calculate the long-wavelength limit of
$\cT_\spe^{-1*}F_{\spe 0}$ to first order in $\epsilon_\spe$. Inverting
\eq{eq:totalchangecoorfirstorder} to first order, and recalling
\eq{eq:totalchangecoorfirstorderCorrectionsLW} and the relations
$\partial_u F_{\spe0 } = -(u/{T}) F_{\spe0 }$, $\partial_\mu
F_{\spe0 } = -(B/{T}) F_{\spe0 }$, one finds 
\eq{eq:pullback_order1_Maxwellian_text}.}

The results of this appendix are also used in equations \eq{eq:T1F1lw}
and \eq{eq:T1F1sw}.

%%%%%%%%%%%%%%%%%%%%%%%%%%%%%%%%%%%%%%%%%%%%%%%%%%%%%%%%%%%%%%%%%%%%%%%%%%%%%%%%
\section{Second-order transformation of the Maxwellian at 
short-wavelengths}
\label{sec:SecondOrderTransfMaxwellianSW}
%%%%%%%%%%%%%%%%%%%%%%%%%%%%%%%%%%%%%%%%%%%%%%%%%%%%%%%%%%%%%%%%%%%%%%%%%%%%%%%%

In this section we derive $[\mathcal{T}^{-1\ast}_{\spe, 2} F_{\spe
  0}]^\sw$. We proceed in a way analogous to that employed in Appendix
G of \cite{CalvoParra2012}. Since $F_{\spe 0}$ is a Maxwellian that
depends only on $\bR$ and $u^2/2 + \mu B(\bR)$, we deduce that
\begin{eqnarray}\label{eq:Tinv2F0}
\fl
\mathcal{T}^{-1\ast}_{\spe, 2} F_{\spe 0}
=
\nonumber\\
\fl\hspace{0.5cm}
 \frac{1}{2B^2} (\bv \times \bun) (\bv \times \bun) : \Bigg [
\nabla_\boldr \nabla_\boldr \ln n_\spe + \left ( \frac{v^2}{2{T}} -
\frac{3}{2} \right ) \nabla_\boldr \nabla_\boldr \ln {T} 
\nonumber\\
\fl\hspace{0.5cm}
-
\frac{v^2}{2{T}^3} \nabla_\boldr {T} \nabla_\boldr {T}
+
\Bigg ( \frac{\nabla_\boldr n_\spe}{n_\spe} + \left ( \frac{v^2}{2{T}}
- \frac{3}{2} \right ) \frac{\nabla_\boldr {T}}{{T}} \Bigg ) \Bigg
( \frac{\nabla_\boldr n_\spe}{n_\spe} 
\nonumber\\
\fl\hspace{0.5cm}
+ \left ( \frac{v^2}{2{T}} -
\frac{3}{2} \right ) \frac{\nabla_\boldr {T}}{{T}} \Bigg ) \Bigg ]
\cT_{\spe,0}^{-1*}F_{\spe 0}
\nonumber\\
\fl\hspace{0.5cm}
+ {\bR}_{02}\cdot \left ( \frac{\nabla_\boldr
n_\spe}{n_\spe} + \left ( \frac{v^2}{2{T}} - \frac{3}{2} \right
) \frac{\nabla_\boldr {T}}{{T}} \right ) \cT_{\spe,0}^{-1*}F_{\spe 0}
\nonumber\\
\fl\hspace{0.5cm}
-
\frac{1}{B} H_{01} (\bv \times \bun) \cdot \Bigg (
\frac{\nabla_\boldr n_\spe}{n_\spe} + \left ( \frac{v^2}{2{T}} -
\frac{5}{2} \right ) \frac{\nabla_\boldr {T}}{{T}} \Bigg )
\frac{\cT_{\spe,0}^{-1*} F_{\spe0}}{{T}}
\nonumber\\
\fl\hspace{0.5cm}
+ \frac{1}{2} H_{01}^2
\frac{\cT_{\spe,0}^{-1*}F_{\spe 0}}{{T}^2} - H_{02}
\frac{\cT_{\spe,0}^{-1*}F_{\spe 0}}{{T}},
\end{eqnarray}
where the functions $\bR_{02}$, $H_{01}$ and $H_{02}$ are given by
\begin{eqnarray}
\fl
\bR = \boldr + \frac{\epsilon_\spe}{B} \bv \times \bun +
\epsilon_\spe^2 \bR_{02}   + O(\epsilon_\spe^3)
\end{eqnarray}
and
\begin{eqnarray}
\fl
\cT_\spe^{-1*}\left(
\frac{u^2}{2} + \mu B(\bR)
\right)
= \frac{v^2}{2} + \epsilon_\spe H_{01}
+ \epsilon_\spe^2 H_{02}+ O(\epsilon_\spe^3).
\end{eqnarray}

Unlike in reference \cite{CalvoParra2012}, we need the short-wavelength
component of \eq{eq:Tinv2F0},
\begin{eqnarray}\label{eq:Tinv2F0sw}
\fl
\left[\mathcal{T}^{-1\ast}_{\spe, 2} F_{\spe 0}\right]^\sw
=
\nonumber\\
\fl\hspace{0.5cm}
+ {\bR}_{02}^\sw\cdot \left ( \frac{\nabla_\boldr
n_\spe}{n_\spe} + \left ( \frac{v^2}{2{T}} - \frac{3}{2} \right
) \frac{\nabla_\boldr {T}}{{T}} \right ) \cT_{\spe,0}^{-1*}F_{\spe 0}
\nonumber\\
\fl\hspace{0.5cm}
-
\frac{1}{B} H_{01}^\sw (\bv \times \bun) \cdot \Bigg (
\frac{\nabla_\boldr n_\spe}{n_\spe} + \left ( \frac{v^2}{2{T}} -
\frac{5}{2} \right ) \frac{\nabla_\boldr {T}}{{T}} \Bigg )
\frac{\cT_{\spe,0}^{-1*} F_{\spe0}}{{T}}
\nonumber\\
\fl\hspace{0.5cm}
+ \frac{1}{2} \left[H_{01}^2\right]^\sw
\frac{\cT_{\spe,0}^{-1*}F_{\spe 0}}{{T}^2} - H_{02}^\sw
\frac{\cT_{\spe,0}^{-1*}F_{\spe 0}}{{T}}.
\end{eqnarray}
The coefficients $\bR_{02}^\sw$ and $H_{01}$ are obtained
following Appendix G of \cite{CalvoParra2012}, easily arriving at
\begin{eqnarray}\label{eq:R02sw}
\fl
\bR_{02}^\sw = \frac{Z_\spe^2\tau_\spe}{B^2}\bun\times
\mathbb{T}_{\spe,0}\nabla_{\bR_\perp/\epsilon_\spe}\tilde\Phi_{\spe 1}^\sw,
\end{eqnarray}
\begin{eqnarray}\label{eq:H01lw}
\fl
H_{01}^\lw = -\frac{Z_\spe}{B}(\bv\times\bun)\cdot
\nabla_\boldr\varphi_0
,
\end{eqnarray}
\begin{eqnarray}\label{eq:H01sw}
  \fl
  H_{01}^\sw = Z_\spe^2\tau_\spe \mathbb{T}_{\spe,0}
\tilde\phi_{\spe 1}^\sw.
\end{eqnarray}
In order to find $H_{02}^\sw$ we recall (G.9) in
reference \cite{CalvoParra2012},
\begin{eqnarray}\label{eq:relationbetweenHamiltonians}
\fl
\cT_{\spe}^*\left(
\frac{v^2}{2} + Z_\spe^2\tau_\spe \epsilon_\spe \varphi
(\boldr,t)
\right)
 = 
 \nonumber\\[5pt]
\fl\hspace{1cm}
\frac{u^2}{2} + \mu B(\bR) 
+ Z_\spe^2\tau_\spe
\epsilon_\spe \langle \phi_\spe \rangle(\bR,\mu,t)
 \nonumber\\[5pt]
\fl\hspace{1cm}
+ Z_\spe^4\tau_\spe^2
\epsilon_\spe^2 \Psi_{\phi,\spe} + Z_\spe^2\tau_\spe \epsilon_\spe^2
\Psi_{\phi B,\spe} + \epsilon_\spe^2 \Psi_{B,\spe}
 \nonumber\\[5pt]
\fl\hspace{1cm}
-
\frac{Z_\spe^2\tau_\spe \epsilon_\spe^2}{B}
\partial_t \Phiwig_\spe
 + O(\epsilon_\spe^3).
\end{eqnarray}
It is almost immediate to see that
\begin{eqnarray}\label{eq:H02sw}
  \fl
  H_{02}^\sw
  = 
  Z_\spe^3\tau_\spe^2\varphi_2^\sw(\boldr,t)
  -Z_\spe^2\tau_\spe
  [\bR_{02}\cdot
\mathbb{T}_{\spe,0}
\nabla_{\bR_\perp/\epsilon_\spe}\langle \phi_{\spe 1}^\sw \rangle]^\sw
  \nonumber\\[5pt]
  \fl\hspace{0.5cm}
  +Z_\spe^2\tau_\spe[\mathbb{T}_{\spe,0}\hat\mu^\sw_{\spe 1}
\mathbb{T}_{\spe,0}
\partial_\mu\langle \phi_{\spe 1}^\sw \rangle
]^\sw
+Z_\spe^2\tau_\spe
\cT_{\spe,0}^{-1*}\hat\mu_{\spe 1}^\lw
\mathbb{T}_{\spe,0}
\partial_\mu\langle \phi_{\spe 1}^\sw \rangle
\nonumber\\[5pt]
  \fl\hspace{0.5cm}
  -Z_\spe^3\tau_\spe^2\mathbb{T}_{\spe,0}\langle \phi_{\spe 2}^\sw \rangle
  -Z_\spe^4\tau_\spe^2\mathbb{T}_{\spe,0}
\Psi_{\phi,\spe} - Z_\spe^2\tau_\spe\mathbb{T}_{\spe,0}\Psi_{\phi B,\spe}
\nonumber\\[5pt]
  \fl\hspace{0.5cm} 
 +
  \frac{Z_\spe^2\tau_\spe}{B}
\mathbb{T}_{\spe,0}
\partial_t\tilde\Phi_\spe^\sw,
\end{eqnarray}
where $\cT_{\spe,0}^{-1*}\hat\mu_{\spe 1}^\lw$ and
$\mathbb{T}_{\spe,0}\hat\mu^\sw_{\spe 1}$ can be found in
\eq{eq:pullbackhatmulw} and \eq{eq:pullbackhatmusw}.

Equation \eq{eq:Tinv2F0sw}, together with \eq{eq:R02sw},
\eq{eq:H01lw}, \eq{eq:H01sw} and \eq{eq:H02sw}, give an explicit
expression for $[\mathcal{T}^{-1\ast}_{\spe, 2} F_{\spe 0}]^\sw$. 

%%%%%%%%%%%%%%%%%%%%%%%%%%%%%%%%%%%%%%%%%%%%%%%%%%%%%%%%%%%%%%%
\section{Computation of the gyroaverage of
  $C_{\spe\spe'}^{(2)\lw}$}
\label{sec:C2nc}
%%%%%%%%%%%%%%%%%%%%%%%%%%%%%%%%%%%%%%%%%%%%%%%%%%%%%%%%%%%%%%%

In this appendix we calculate the
gyroaverage of \eq{eq:C2lw}. First, we can write 
\begin{eqnarray}\label{eq:C2lwgyro}
\fl\Big\langle \cT_{\spe, 0}^{*}C_{\spe \spe'}^{(2)\lw}\Big\rangle =
\cT_{\spe, 0}^{*} C_{\sigma \sigma^\prime}
 \Big[\cT_{\spe, 0}^{-1*}\left\langle F_{\sigma 2}^\lw \right\rangle +
\cT_{\spe, 0}^{-1*}\left\langle 
\cT_{\spe, 0}^{*} [\cT_{\sigma,1}^{-1 *} F_{\sigma 1}^\lw]^\lw\right\rangle 
\nonumber\\\hspace{0.5cm}
\fl
 +
\cT_{\spe, 0}^{-1*}\left\langle \cT_{\spe, 0}^{*}
[\cT_{\sigma,2}^{-1 *} F_{\spe 0}]^\lw\right\rangle ,
\cT^{-1*}_{\spe',0}F_{\spe' 0} \Big] 
+\left(\frac{Z_\spe\tau_\spe}{Z_{\spe'}\tau_{\spe'}}\right)^2
\cT_{\spe, 0}^{*}C_{\sigma \sigma^\prime}
\Big[
\cT^{-1*}_{\spe,0}F_{\spe 0},
\nonumber\\\hspace{0.5cm}
\fl
\cT_{\spe^\prime, 0}^{-1*}
 \left\langle F_{\sigma' 2}^\lw\right\rangle +
\cT_{\spe^\prime, 0}^{-1*}
\left\langle
\cT_{\spe^\prime, 0}^{*}[\cT_{\sigma',1}^{-1 *} F_{\sigma' 1}^\lw]^\lw\right\rangle +
\cT_{\spe^\prime, 0}^{-1*}\left\langle\cT_{\spe^\prime, 0}^{*}
[\cT_{\sigma',2}^{-1 *} F_{\spe' 0}]^\lw\right\rangle
\Big] \nonumber\\\hspace{0.5cm}
\fl +
\frac{Z_\spe\tau_\spe}{Z_{\spe'}\tau_{\spe'}}
\left\langle
\cT_{\spe, 0}^{*}C_{\sigma\sigma^\prime}
\left[\cT^{-1*}_{\spe,0} F_{\sigma 1}^\lw
+ [\cT_{\sigma,1}^{-1 *} F_{\sigma 0}]^\lw,
\cT^{-1*}_{\spe',0}F_{\sigma^\prime 1}^\lw +
[\cT_{\sigma',1}^{-1 *} F_{\sigma' 0}]^\lw
\right]
\right\rangle
\nonumber\\\hspace{0.5cm}
\fl +
\frac{Z_\spe\tau_\spe}{Z_{\spe'}\tau_{\spe'}}
\Bigg
\langle
\cT_{\spe, 0}^{*} \Bigg[C_{\sigma\sigma^\prime} \Bigg[ 
\modTinv F_{\spe 1}^\sw -
\frac{Z_\spe^2\tau_\spe
}{{T}}
\modTinv \phiwig_{\spe 1}^\sw\cT_{\spe,0}^{-1 *}F_{\spe 0}
, 
\modTinvprime F_{\spe' 1}^\sw \nonumber\\\hspace{0.5cm}
\fl
-
\frac{Z_{\spe'}^2\tau_{\spe'} 
}{{T}}
\modTinvprime\phiwig_{\spe' 1}^\sw\cT_{\spe',0}^{-1 *}F_{\spe' 0}
\Bigg] \Bigg]^\lw
\Bigg
\rangle,
\end{eqnarray}
where we have used that $\left\langle
\cT_{\spe, 0}^{*}[\cT_{\sigma,1}^{-1 *} F_{\sigma
    1}^\sw]^\lw\right\rangle = 0$. Here,
\begin{eqnarray}\label{eq:T1F1lwAveraged}
\fl \left\langle
\cT_{\spe, 0}^{*}[\cT_{\sigma,1}^{-1 *} F_{\spe 1}^\lw]^\lw
\right\rangle = 
\mu\bun\cdot\nabla_\bR\times\bun
\left(
\partial_u
-\frac{u}{B} 
\partial_\mu
\right)
F_{\spe 1}^\lw,
\end{eqnarray}
and 
\begin{eqnarray}\label{gyroaverageT2F0}
\fl\left\langle\cT_{\spe, 0}^{*} [\cT_{\sigma,2}^{-1 *} F_{\spe 0}]^\lw\right\rangle
 =\nonumber\\
\fl\hspace{1cm} \frac{\mu}{2B} (\matI - \bun \bun) : \Bigg [
\nabla_\bR \nabla_\bR \ln n_\spe + \left ( \frac{u^2/2 + \mu B}{{T}} -
\frac{3}{2} \right ) \nabla_\bR \nabla_\bR \ln {T} \Bigg ] F_{\spe 0}
\nonumber\\
\fl\hspace{1cm} 
- \frac{\mu}{B} \frac{Z_\spe}{{T}^2}
\nabla_\bR \varphi_0 \cdot \nabla_\bR {T} F_{\spe 0} -
\frac{\mu}{2B} \frac{u^2/2 + \mu B}{{T}^3} |\nabla_\bR {T}|^2
F_{\spe 0} \nonumber\\
\fl\hspace{1cm} 
+ \frac{\mu}{2B} \Bigg |
\frac{\nabla_\bR n_\spe}{n_\spe} + \frac{Z_\spe \nabla_\bR
\varphi_0}{{T}} + \left ( \frac{u^2/2 + \mu B}{{T}} - \frac{3}{2}
\right ) \frac{\nabla_\bR {T}}{{T}} \Bigg |^2 F_{\spe 0}
\nonumber\\
\fl\hspace{1cm} 
- \frac{\mu}{2B^2} \nabla_{\bR_\bot} B \cdot \left (
\frac{\nabla_\bR n_\spe}{n_\spe} + \frac{Z_\spe \nabla_\bR
\varphi_0}{{T}} + \left ( \frac{u^2/2 + \mu B}{{T}} - \frac{3}{2}
\right ) \frac{\nabla_\bR {T}}{{T}} \right ) F_{\spe 0}
\nonumber\\
\fl\hspace{1cm} 
+ \frac{Z_\spe^4\tau_\spe^2}{2{T}^2}
\left[ \left \langle
(\phiwig_{\spe 1}^\sw)^2 \right \rangle \right]^\lw F_{\spe
0} + \frac{1}{{T}} \Bigg [ - \frac{Z_\spe^2}{2B^2}
|\nabla_\bR \varphi_0|^2 \nonumber\\
\fl\hspace{1cm} 
- \frac{Z_\spe^4\tau_\spe^2}{2B}
\partial_\mu \left[ \left
\langle (\phiwig_{\spe 1}^\sw)^2 \right \rangle \right]^\lw
- \frac{3 Z_\spe \mu}{2 B^2} \nabla_{\bR_\bot} B \cdot
\nabla_\bR \varphi_0
\nonumber\\
\fl\hspace{1cm} 
 - \frac{Z_\spe u^2}{B^2} \bun
\cdot \nabla_\bR \bun \cdot \nabla_\bR \varphi_0  + \Psi_{B, \spe}
\nonumber\\
\fl\hspace{1cm} 
+ 
\frac{Z_\spe
\mu}{B} (\matI - \bun \bun): \nabla_\bR \nabla_\bR \varphi_0 \Bigg
] F_{\spe 0}.
\end{eqnarray}
This last result has been obtained by gyroaveraging \eq{T2F0}.

%%%%%%%%%%%%%%%%%%%%%%%%%%%%%%%%%%%%%%%%%%%%%%%%%%%%%%%%%%%%%%%%%%%%%%%%%%%%%
\section*{References}
%%%%%%%%%%%%%%%%%%%%%%%%%%%%%%%%%%%%%%%%%%%%%%%%%%%%%%%%%%%%%%%%%%%%%%%%%%%%%

\end{document}